\documentclass[11pt,a4paper]{article}
\pdfoutput=1
\usepackage[usenames, dvipsnames]{color}
\usepackage[compat=1.1.0]{tikz-feynman}
\usepackage{jheppub}
\usepackage{amssymb,amsmath}
\usepackage{url,booktabs}
\usepackage{slashed}
\usepackage{multirow}
\def\be{\begin{equation}}
\def\ee{\end{equation}}
\def\bea{\begin{eqnarray}}
\def\eea{\end{eqnarray}}

\graphicspath{{Plots/}}
\relax

\title{LHC-friendly minimal freeze-in models}
\author[a]{G. B\'elanger}
\author[b]{N. Desai}
\author[c]{A. Goudelis}
\author[d]{J. Harz}
\author[e]{A. Lessa}
\author[f]{J.M. No}
\author[g]{A. Pukhov}
\author[h]{S. Sekmen}
\author[i]{D. Sengupta}
\author[a,f]{B. Zaldivar}
\author[j,k]{J. Zurita}
\affiliation[a]{LAPTh, Univ. Grenoble Alpes, USMB, CNRS,  74940 Annecy, France}
\affiliation[b]{Laboratoire Charles Coulomb (L2C) and Laboratoire Univers et Particules de Montpellier (LUPM), CNRS-Universit\'e de Montpellier, France}
\affiliation[c]{Sorbonne Universit\'e, CNRS, Laboratoire de Physique Th\'eorique et Hautes \'Energies, LPTHE, F-75252 Paris, France and Sorbonne Universit\'es, Institut Lagrange de Paris (ILP), 98 bis Boulevard Arago, 75014 Paris, France
}
\affiliation[d]{Physik  Department  T70,  James-Franck-Stra{\ss}e, Technische  Universit{\"a}t  M{\"u}nchen, 85748  Garching,  Germany}
\affiliation[e]{Centro de Ci$\hat{e}$ncias Naturais e Humanas, Universidade Federal do ABC, Santo Andr\'e, 09210-580 SP, Brazil}
\affiliation[f]{Departamento de Fisica Teorica and Instituto de Fisica Teorica, IFT-UAM/CSIC,
Cantoblanco, 28049, Madrid, Spain.}
\affiliation[g]{Skobeltsyn Institute of Nuclear Physics, Moscow State University, Moscow 119992, Russia.}
\affiliation[h]{Kyungpook National University, Physics Dept., Daegu, South Korea.}
\affiliation[i]{Department of Physics and Astronomy, Michigan State University, East Lansing, MI, U.S.A}
\affiliation[j]{Institute for Nuclear Physics (IKP), Karlsruhe Institute of Technology, Hermann-von-Helmholtz-Platz 1, D-76344 Eggenstein-Leopoldshafen, Germany}
\affiliation[k]{Institute for Theoretical Particle Physics (TTP), Karlsruhe Institute of Technology, Engesserstra{\ss}e 7, D-76128 Karlsruhe, Germany} 
\emailAdd{belanger@lapth.cnrs.fr}
\emailAdd{nishita.desai@umontpellier.fr}
\emailAdd{andreas.goudelis@lpthe.jussieu.fr}
\emailAdd{julia.harz@tum.de}
\emailAdd{andre.lessa@ufabc.edu.br}
\emailAdd{josemiguel.no@uam.es}
\emailAdd{pukhov@lapp.in2p3.fr}
\emailAdd{ssekmen@cern.ch}
\emailAdd{dipan@pa.msu.edu}
\emailAdd{bryan.zaldivarm@uam.es}
\emailAdd{jose.zurita@kit.edu}
\abstract{We propose simple freeze-in models where the observed dark matter abundance is explained via the decay of an electrically charged and/or coloured parent particle into Feebly Interacting Massive Particles (FIMP). The parent particle is long-lived and yields a wide variety of LHC signatures depending on its lifetime and quantum numbers. We assess the current constraints and future high luminosity reach of these scenarios at the LHC from searches for heavy stable charged particles, 
disappearing tracks, displaced vertices and displaced leptons. We show that the LHC constitutes a powerful probe of freeze-in dark matter 
and can further provide interesting insights on the validity of vanilla baryogenesis and leptogenesis scenarios.}

\begin{document}
\begin{flushright}
TUM-HEP-1173-18 \\ TTP18-040 \\ IFT-UAM/CSIC-18-113 \\ LAPTH-046/18
\end{flushright}
\maketitle
\notoc

\section{Introduction}
\label{sec:intro}

The frozen-out Weakly Interacting Massive Particle (WIMP) picture has dominated the dark matter (DM) model building scene for the last 30 years. Among its attractive features is the fact that it relates the measured DM abundance~\cite{Ade:2015xua} to a particle with $SU(2)_L$ weak couplings and weak scale masses ($m_{\chi} \sim {\cal O}(100-1000)$ GeV), an energy at which physics beyond the Standard Model (SM) may manifest itself. However, the lack of any conclusive and undisputed signals at colliders \cite{Sirunyan:2017jix,Aaboud:2017phn} as well as direct detection~\cite{Cui:2017nnn,Amole:2017dex,Akerib:2017kat,Aprile:2018dbl} and indirect detection~\cite{Ahnen:2016qkx,Abdallah:2016ygi} experiments has put this \emph{beau ideal} under siege. Indeed, most of the simplest models of WIMP-type dark matter~\cite{Cirelli:2005uq,Abdallah:2015ter,Boveia:2016mrp,DeSimone:2016fbz} are currently under substantial tension with null experimental evidence, and many of the surviving scenarios are expected to be probed in the near future or, eventually, at a future facility (see e.g~\cite{Arcadi:2017kky,Low:2014cba,Cirelli:2014dsa,Ostdiek:2015aga,Mahbubani:2017gjh,Krall:2017xij,Gori:2014oua}).

While evading these constraints within the freeze-out framework will, most likely, remain possible over the next few decades either by increasing the DM mass or by rendering the ``dark sector'' more complex~\cite{Griest:1990kh,DAgnolo:2017dbv,Garny:2017rxs,DAgnolo:2018wcn,Evans:2017kti}, another route is to instead consider alternative dark matter production mechanisms in the early Universe \cite{Baer:2014eja}. Among the various proposed mechanisms, the freeze-in 
picture~\cite{McDonald:2001vt,Hall:2009bx} is a particularly interesting possibility, since it constitutes a relatively simple scenario to explain the observed DM abundance in the Universe that can be invoked in numerous well-motivated extensions of the SM (see e.g.~\cite{Covi:1999ty, Cheung:2011nn,Asaka:2005cn,Goudelis:2018xqi,Garny:2018ali,Heeba:2018wtf}). The main idea behind freeze-in is that dark matter production is driven by processes involving tiny couplings, $y \sim { \cal O}(10^{-13} - 10^{-7})$,  which never allow it to attain chemical equilibrium with the SM thermal bath during the cosmic evolution. Given the extremely weak interactions between the two sectors, freeze-in dark matter candidates are, in general, compatible with the current null results from direct and indirect searches and the phenomenology of such Feebly Interacting Massive Particle (FIMP) dark matter is more challenging than that of WIMPs~\cite{Bernal:2017kxu}. Nonetheless, in the case where the interaction between DM and the SM is mediated by light enough particles, conventional direct detection experiments could test substantial parts of the freeze-in parameter space \cite{Essig:2011nj,Hambye:2018dpi,Knapen:2017xzo}, whereas some freeze-in scenarios could even give rise to observable signals in indirect dark matter searches \cite{Heikinheimo:2018duk}. Lastly, collider signatures can arise quite straightforwardly in a variety of freeze-in models, motivating extremely interesting searches at the LHC \cite{Hall:2009bx,Co:2015pka,Calibbi:2018fqf}. In particular, assuming that DM production occurs via the decays of a new particle  $Y$ with sizable couplings to the SM along with some visible (i.e: SM) daughter particle $X$,  $Y \to {\rm DM} + X$, then: {\it (i)} since the particle $Y$ was in equilibrium with the SM in the early Universe, it can potentially be copiously produced at colliders. {\it (ii)} due to the smallness of its coupling to DM, $Y$ should be fairly long-lived and give rise signatures in the LHC detectors that are different from prompty decaying particles.

For similar reasons as those that have lead to a critical reevaluation of the WIMP picture as the dominant explanation for the abundance of DM in the Universe, namely the lack of conclusive signals in prompt LHC searches for physics beyond the SM, in the last few years there has been an upsurge of interest in searches for Long-Lived Particles (LLPs) at the LHC and beyond (see e.g.~\cite{Curtin:2018mvb,LLP_Report}).
While LLP searches have been conducted at hadron colliders already since Tevatron times \cite{Abazov:2006as,Abazov:2007ht,Abazov:2008zm,Abazov:2008qu,Abazov:2009ik,Abazov:2011pf,Abazov:2012ina,Aaltonen:2009kea,Abe:1998ee}, the results have typically been presented in terms of concrete SUSY incarnations, which render their reinterpretation complicated, if not impossible. It is indeed true that the LLP community has recently put a substantial amount of effort into providing ``model-independent" recipes to recast the existing searches for arbitrary models. However, current prescriptions are still not exhaustive and can only be applied to models with some similarity to the benchmark model in terms of which the results are presented (see e.g.~\cite{LLP_Report} for more details on this matter).

Likewise, although numerous freeze-in models have been presented in the literature,  
and although some facets of the collider phenomenology of such models have been pointed out \cite{Co:2015pka,Hessler:2016kwm,Ghosh:2017vhe,Calibbi:2018fqf}, a systematic exploration of the collider signatures predicted by freeze-in scenarios is largely wanting. 
 
It is still unclear which regions of parameter space of freeze-in models can be probed by different LHC LLP searches, how these searches could be optimised in order to target freeze-in models and what would be the optimal way to present experimental results in order to render them as widely applicable as possible. In this spirit, any attempt to build -- simple, yet fairly generic -- freeze-in models involving LLPs, to extract their predictions, to recast LLPs searches in terms of these models and to propose avenues for the further optimisation of these searches provides valuable information for theorists and experimentalists alike.

In this paper we 
propose simple and fully consistent 
freeze-in dark matter models that can be probed at the LHC, generalising our preliminary results from~\cite{Brooijmans:2018xbu}. In all 
models, the dark matter sector can be populated via the decay of an electromagnetic- or colour-charged particle into DM along with a (collider-visible) SM particle. The decay, being non-prompt, gives rise to a multitude of LHC signatures, depending on the $SU(3)_c \times SU(2)_L \times U(1)_Y$ transformation properties and the lifetime of the parent particle. We recast LHC searches based on these signatures, demonstrate their capacity to constrain different regions of the cosmologically favoured parameter space of our models and highlight their complementarity. We moreover provide projections for the High Luminosity LHC (HL-LHC) sensitivity and discuss ways to optimise relevant searches. In addition to the LHC results we also highlight the interesting interplay between freeze-in dark matter, LHC searches for long-lived particles and the origin of the matter-antimatter asymmetry of the Universe.

The paper is structured as follows: in Section \ref{sec:mod} we present our models and estimate the constraints they are subject to from particle physics experiments other than LHC LLP searches. In Section \ref{sec:cosmo} we recall some key features of the freeze-in dark matter production mechanism, discuss cosmological constraints on our models and point out that the observation of a signal in agreement with freeze-in dark matter could falsify simple electroweak baryogenesis and leptogenesis explanations for the origin of the matter-antimatter asymmetry of the Universe. In Section \ref{sec:collider} we turn our attention to LHC LLP searches that constrain our models, namely searches for heavy stable charged particles (HSCP), $R$-hadrons, displaced vertices (DV), displaced lepton searches (DL) and disappearing tracks (DT). We recast all analyses using the latest 8 TeV and 13 TeV LHC data and present projections to the end-of-lifetime HL-LHC reach. While such projections are simplistic and need to be taken with a grain of salt, they elucidate the unique capabilities of the LHC to probe a large region of the viable parameter space and how different searches can test different particle physics and cosmological scenarios. We reserve Section~\ref{sec:conclu} for our conclusions and outlook. In Appendix~\ref{app:HSCP} we discuss some more technical issues concerning the reinterpretation of searches for HSCPs in terms of scenarios involving heavy charged particles with a macroscopic but finite lifetime, and in Appendix~\ref{app:stat} we discuss the statistical procedure for deriving the limits from the LHC analyses. 
\section{Minimal freeze-in dark matter scenarios}
\label{sec:mod}

\subsection{Preliminaries}
\label{sec:preliminaries}

A necessary ingredient in any freeze-in DM model is the existence of some particle that has a negligible initial abundance and interacts feebly (via a coupling $y_\chi$) with the thermal bath, 
thus being thermally decoupled from the latter. In what follows we will assume that the only FIMP state is the DM particle $\chi$ itself. In order to avoid thermalization of $\chi$ via SM gauge interactions leading to the standard DM freeze-out scenario, $\chi$ must be a SM gauge singlet. In addition, we will guarantee the DM stability by imposing a discrete $\mathbf{Z}_2$ symmetry under which $\chi$ is odd whereas the SM particles are even. 

Our aim is to construct consistent models which are \textit{minimal}, in the sense of introducing the least number of exotic fields that are sufficient for successful DM freeze-in but which, at the same time, 
allow us to obtain (in principle) testable collider signatures 
and can accurately capture the DM phenomenology of more complicated theories beyond the SM. The simplest option for a freeze-in DM scenario would be to add just the $\chi$ field to the SM, but the only renormalizable operator between the SM and the dark sector respecting all symmetries is the Higgs portal operator $y_\chi H^2 \chi^2$, where $\chi$ is a spin-0 field. Then in the absence of other couplings, the rate of DM production at colliders would be proportional to the freeze-in coupling $y_\chi$ and yield no observable signature\footnote{The HL-LHC is expected to produce $\sim 10^8$ Higgs bosons. 
Typical freeze-in values for $y_\chi$ lie in the ${\cal O}(10^{-13} - 10^{-7})$ range, would yield less than $10^{-4}$ DM events at the HL-LHC in our scenario.}. 

The next possible step is to add a new BSM field $Y$, which can be either even or odd under the $\mathbf{Z}_2$ symmetry. If the new field is $\mathbf{Z}_2$-even then the construction is reminiscent of the so-called ``simplified models of dark matter'' 
at the LHC (see e.g.~\cite{Abdallah:2015ter,Boveia:2016mrp,DeSimone:2016fbz} and references therein), where $Y$ would act as a mediator 
to the dark sector. In order to produce $Y$ at colliders it requires some sizable couplings to the SM fields. The smallness of the $Y-\chi$ coupling required for DM freeze-in to take place (see Section~\ref{sec:FreezeIn} for a detailed discussion) would yield a tiny branching fraction of $Y$ into the dark sector, leaving the $Y$ resonant decay into SM particles as its only detectable final state at the LHC.

We are thus left with the option of adding a new BSM field $Y$ which is $\mathbf{Z}_2$-odd and whose interactions with the DM particle $\chi$ are of the form 
\be
\label{eq:fiOperator}
y_{\chi}\, Y\, X_{\rm SM}\, \chi
\ee
where $X_{\rm SM}$ labels any SM field. $Y$ then carries the gauge quantum numbers of the corresponding \emph{anti}-SM field. Note that in this construction no coupling beyond $y_\chi$ is required, and the collider phenomenology will be mainly driven by the known SM gauge couplings that $Y$ inherits, while the relic density can follow the DM freeze-in history for an appropriate value of $y_\chi$.

The classification of different possible SM gauge quantum numbers for the fields $X_{\rm SM}$ and $Y$ has been performed in the context of coannihilation scenarios, see Table 2 of~\cite{Baker:2015qna}.  Note that $X_{\rm SM}$ cannot be a gauge boson: while allowed by Lorentz invariance, gauge symmetry would force $Y$ and $\chi$ to be the same multiplet and $y_{\chi}$ would be a SM gauge coupling yielding the thermalization of $\chi $ in the early Universe. Hence $X_{\rm SM}$ is either the Higgs doublet or a fermion (left- or right-handed quark or lepton), and for every possible choice of $X_{\rm SM}$ there are two different spin assignments for the dark sector fields. For concreteness, we will pick $Y$ to be a fermion and $\chi$ to be scalar\footnote{Since, as we will see later on, the DM production at the LHC will occur via the $YY$ final state, fermions will have slightly larger production cross-sections than scalars.}.

Models featuring left-handed weak doublets are characterised by a more varied range of signatures due to the presence of additional degrees of freedom. Whereas these signatures can actually drive the collider phenomenology of such scenarios, they are fairly model-dependent and can obscure the importance of more generic freeze-in-specific signals. Motivated by this, and in the spirit of keeping our models as minimal as possible, we restrict ourselves to the simplest cases where $X_{\rm SM}$ carries no $SU(2)_L$ charge. We will focus on two benchmarks characterised by different $SU(3)_c$ transformation properties of $Y$, meaning that $Y$ will have the same gauge quantum numbers as a right-handed charged lepton or a right-handed up-type quark respectively. 

Before continuing, let us remark that most of the existing collider studies of freeze-in models have instead focused on $Y$ being a weak 
doublet\footnote{See Ref.~\cite{Evans:2016zau} for an exception to this, mostly focused on $\chi$ being a fermion (a gravitino in Gauge-Mediated Supersymmetry Breaking scenarios) and $Y$ being a scalar lepton (a right-handed stau), but also considering the case of a scalar $\chi$ and a fermionic, right-handed $Y$.}. The case in which $X_{\rm SM}$ is the SM Higgs doublet and both $Y$ and $\chi$ are fermions has been studied with $Y$ and $\chi$ being the Higgsino and axino respectively~\cite{Co:2015pka}, as well as in the context of the singlet-doublet DM scenario~\cite{Calibbi:2018fqf}. In addition, the 
study of Ref.~\cite{Hessler:2016kwm} employs the operator in Eq.~\eqref{eq:fiOperator} assuming $X_{\rm SM}$ to be a left-handed lepton doublet, $\chi$ a fermion and $Y$ a scalar doublet. With the previous remarks in mind, we can now present the concrete freeze-in models that we will study in what follows.

\subsection{Minimal freeze-in models with a charged parent}
\label{subsec:models}

 We augment the SM by a real scalar DM candidate $s$ that is neutral under $SU(3)_c \times SU(2)_L \times U(1)_Y$, along with a vector-like fermion $F$, which corresponds to $Y=F$ and $\chi=s$ 
 in the notation of Section~\ref{sec:preliminaries}. As we already mentioned, in all cases we take $F$ to be an $SU(2)$ singlet, whereas both $s$ and $F$ are taken to be odd under a $\mathbf{Z}_2$ symmetry which, by choosing $m_s < m_{F}$, stabilizes the DM candidate. The SM particles are taken to transform trivially under the same discrete symmetry.
 
We couple the DM candidate to the SM through Yukawa-type terms involving the left-handed component of the vector-like fermion and the SM right-handed fermions: up-type quarks, down-type quarks and charged leptons. The Lagrangian for all three models can be succinctly written as
\begin{align}
\label{eq:lag}
{\cal{L}} & = {\cal{L}}_{\rm SM} + \partial_\mu s ~ \partial^\mu s  - \frac{\mu_s^2}{2} s^2 + \frac{\lambda_s}{4} s^4 + \lambda_{sh} s^2 \left(H^\dagger H\right) \\ \nonumber
& + \bar{F} \left(i\slashed{D}\right) F  - m_{F} \bar{F} F - 
\sum_{f} y_{s}^{f} \left(s \bar{F} \left( \frac{1+\gamma^5}{2} \right) f + {\rm{h.c.}} \right),
\end{align}
where $f = \lbrace e, \mu, \tau \rbrace$, $\lbrace u, c, t \rbrace$ or $\lbrace d, s, b \rbrace$, depending on the $SU(3)_c \times U(1)_Y$ transformation properties of $F$\footnote{The same Lagrangian has been employed in DM freeze-out phenomenological studies, see e.g~\cite{Giacchino:2015hvk,Colucci:2018vxz,Colucci:2018qml}.}.
The three models are each described by a set of seven free parameters. We choose these parameters to be
\begin{equation}\label{eq:parameters}
m_s, m_F, \lambda_{sh}, \lambda_{s}, \lbrace y_{s}^{f} \rbrace \ .
\end{equation}
The first four are common to the three models and correspond to the DM mass, the vector-like fermion mass, the DM-Higgs quartic coupling and the DM quartic self-coupling respectively. Since the DM self-coupling, $\lambda_{s}$, is irrelevant for our purposes, we set it to zero\footnote{For cases in which such interactions can become important see e.g. \cite{Bernal:2015ova}.}. Note that the dark scalar mass is related to the $\mu_s$ parameter entering Eq.\eqref{eq:lag} through
\begin{equation}
\mu_s^2 = m_s^2 + \lambda_{sh} v^2 \ .
\end{equation}
Although the Higgs portal term could contribute to the freeze-in dynamics, it is not the main focus of this paper and throughout this work we will 
set $\lambda_{sh}=0$\footnote{For $\lambda_{sh}$ to be irrelevant to the present analysis, $\lambda_{sh} \ll 10^{-11}$ is needed~\cite{Yaguna:2011qn}. Note however that this operator would be absent had we chosen $Y$ to be a scalar and $\chi$ to be a fermion.}.
The last three parameters $\lbrace y_{s}^{f} \rbrace$) in Eq.~\eqref{eq:parameters} determine the interaction strength of the DM particle to the visible sector, and the freeze-in mechanism forces them to be small. Note that since we are assume no specific flavor protection mechanism, our setup naturally leads to flavor violation, both in the lepton and quark sector. These effects are naturally suppressed due to the size of the freeze-in coupling, but we will comment on them in the context of each specific model in the next subsections. 

In order to perform our phenomenological analysis, we have implemented the three models described by the Lagrangian of Eq.\eqref{eq:lag} in the {\tt FeynRules} package \cite{Alloul:2013bka} and exported them in {\tt UFO} \cite{Degrande:2011ua} and {\tt CalcHEP} \cite{Belyaev:2012qa} file format for use with {\sc MadGraph5\_aMC@NLO} \cite{Alwall:2014hca} and {\tt micrOMEGAs 5} \cite{Belanger:2018ccd}. The three cases (couplings to leptons, up- and down-type quarks) have been implemented separately and the corresponding model files can be found in the {\tt FeynRules} Model Database \cite{FRmodeldatabase}, or directly in~\cite{FICPLHC}.

Before discussing the cosmology and LHC phenomenology of our models, we will briefly comment on potential additional constraints in each specific case.

\subsubsection{Coupling to leptons}
\label{section:leptonmodel}

\begin{figure}
\centering
\begin{tikzpicture}
\begin{feynman}
  \node [blob] (a);
  \vertex [above left=of a] (i1) {$p$};
  \vertex [below left=of a] (i2) {$p$};
  \vertex [right=of a] (b);
  \node [above right=of b,circle,fill,inner sep=1pt] (E1);
  \node [below right=of b,circle,fill,inner sep=1pt] (E2);
  \vertex [above right=of E1] (l1) {$l$};
  \vertex [below=of l1] (s1) {$s$};
  \vertex [below right=of E2] (l2) {$l$};
  \vertex [above=of l2] (s2) {$s$};

\diagram*{
   (i1) -- [double] (a) -- [double] (i2),
   (E1) -- [fermion,edge label'={$F$}] (b) -- [fermion,edge label'={$F$}] (E2),
   (s1) -- [scalar] (E1) -- [anti fermion] (l1), 
   (s2) -- [scalar] (E2) -- [fermion] (l2),
   (a) -- [photon,edge label'={$\gamma,Z$}] (b),
};
\end{feynman}
\end{tikzpicture}
\caption{\it Diagram for the main production and decay process of $F$
at the LHC, in the leptonic model.}
\label{fig:LHC_production}
\end{figure}
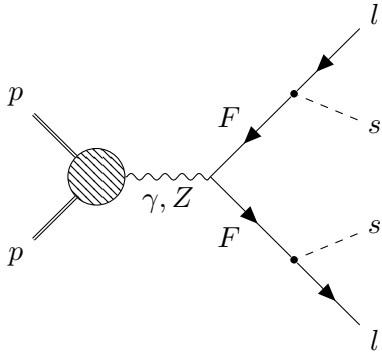

 In this variant of the model we take $f \equiv \ell = \lbrace e, \mu, \tau \rbrace$, which implies that $F$ transforms as $(\mathbf{1},\mathbf{-1})$ under $SU(3)_c \times U(1)_Y$. Since DM communicates with the SM through a Yukawa-type interaction, we need to pick a flavor structure for the interaction terms in Eq.~\eqref{eq:lag}. Since several of the LHC searches considered in section~\ref{subsec:DL} require displaced decays (i.e: within the detector but away from the primary vertex) to electrons and/or muons, we will only consider couplings to the first two generation leptons\footnote{It is also conceivable to couple the $s$ and $F$ fields to $\tau$ leptons. An analysis of a model with interactions to all $e, \mu$ and $\tau$ can be found in~\cite{Evans:2016zau}, with the key difference that the dark sector particles have the opposite spins, namely $\chi$ is a fermion (gravitino in SUSY) and $Y$ corresponds to a scalar lepton, $\tilde{\tau}$.}.  The LHC signature of the model, illustrated in Figure~\ref{fig:LHC_production}, is the Drell-Yan pair-production of $F$ followed by the $F \to s \ell$ decay, which can be displaced or even take place outside the detector. 

LEP2 constraints are relevant and we expect a bound on $m_F > 104$ GeV, namely half the maximum-center of mass energy. However, there are some loopholes in this statement: taking into account the decay length $c \tau$ of $F$ we actually obtain  
\begin{itemize}
\item $m_F > 102$ GeV for $0.3\, \mathrm{m} \lesssim c\tau \lesssim\, 3$ mm.
\item $m_F > 100$ GeV for $3\, \mathrm{m} \lesssim c\tau \lesssim 300\, \mathrm{m}$ (assuming 100\% decay into muons).
\item  $m_F > 100$ GeV for $0.3 \,\mathrm{m} \lesssim c\tau \lesssim 300$ mm (assuming 100\% decay into electrons).
\end{itemize}
We have extracted these bounds from the two lower panels of Figure 7 of Ref.~\cite{Abbiendi:2005gc}. As stressed in the recent literature, these loopholes from LEP reach can be probed at the LHC with dedicated searches~\cite{Egana-Ugrinovic:2018roi}. As for current collider constraints the current searches for sleptons exclude masses up to 500 GeV~\cite{Aaboud:2018jiw} if the neutralino mass is below 300 GeV. Reinterpreting these searches for long-lived particles would lead to a degradation in sensitivity. Indeed, the ATLAS collaboration has carried out the analysis for the case of colored fermions~\cite{ATLAS:2018yey}, finding that the prompt searches do not compete with the dedicated long-lived ones when $c \tau \gtrsim 0.5$ cm. Thus we conclude we are safe from constraints from prompt searches.

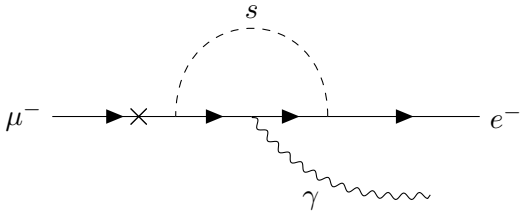
\begin{figure}[thb]
\centering
\begin{tikzpicture} \begin{feynman}
\vertex (a1) {\( \mu^{-} \)}; \vertex[right=2cm of a1] (a2); \vertex[right=1cm of a2] (a3); \vertex[right=1cm of a3] (a4); \vertex[right=2cm of a4] (a5) {\(e^{-} \)};
\vertex[below=1.5em of a5] (c1) {\( \)}; \vertex[below=1.5cm of a5] (c3) {\( \)}; \vertex at ($(c1)!0.5!(c3) - (1cm, 0)$) (c2) ;
\diagram* { {[edges=fermion]
(a1) -- [insertion=0.7] (a2) -- (a3) -- (a4) -- (a5), },
(c3) -- [fermion, out=180, in=-60, opacity=0.0001] (c2) -- [fermion, out=60, in=180, opacity=0.0001] (c1), (a3) -- [photon, bend right, edge label'=\( \gamma \)] (c2),
(a4) -- [scalar, out=90, in=90, looseness=2.0, edge label'=\(s\)] (a2)
};
\end{feynman} \end{tikzpicture}
\caption{Representative Feynman diagram for $\mu \to e \gamma$.}
\label{fig:LFVdiagrams}
\end{figure}

Regarding indirect constraints, we first stress that for vector-like fermions which do not mix with the SM fermions and are $SU(2)_L$ singlets, there are no relevant contributions to the electroweak precision observables (see e.g. the discussion in~\cite{Ellis:2014dza}), which we can therefore safely ignore.
Secondly, for a sufficiently light $s$, the last term in Eq.~\eqref{eq:lag} can introduce an additional muon decay channel, $\mu \rightarrow ess$. We have numerically checked that the corresponding width is well below the current experimental precision of muon lifetime measurements~\cite{Tanabashi:2018oca}.
Lastly, as we do not assume any flavour protection, contributions to lepton flavour violating (LFV) processes can arise. We should, hence, consider constraints from LFV experiments, such as $Br(\mu \to e \gamma) < 4.2 \times 10^{-13}~(90\%~ C.L.)$ \cite{TheMEG:2016wtm}, $R^{Ti}_{\mu^- e^-} < 7 \times 10^{-13}~(90\%~ C.L.)$ \cite{Bertl:2006up}, and $Br(\mu^+ \to e^+ e^- e^+) < 1.0 \times 10^{-12}~(90\%~ C.L.)$ \cite{Bellgardt:1988qe}. As our model contributes dominantly to the dipole operator and not to the four-fermion operator, $\mu \to e \gamma$ provides the most stringent limit, the leading contribution of which is via de diagram shown in Figure~\ref{fig:LFVdiagrams}. In order to approximate the corresponding branching ratio we can adjust the general expressions for the process $f_1 \to f_2 \gamma$ presented in Ref.~\cite{Lavoura:2003xp}. Assuming $m_s \ll m_F$ (a regime which, as we will see, will be particularly interesting for us in the following) and ignoring the tiny contribution of $\mu \rightarrow ess$ to the total muon decay width, we obtain
\begin{align}
    Br(\mu \to e \gamma) \approx \frac{2 v^4 (y^e_s)^2 (y^\mu_s)^2}{3 m_F^4 (16 \pi)^2} \, .
\end{align}
For typical freeze-in-motivated parameter choices we find $Br(\mu \to e \gamma) \sim \mathcal{O}(10^{-46})$, which is far beyond the sensitivity of current and future experiments.

\subsubsection{Coupling to quarks}

A simple model of FIMP DM coupled to quarks can be constructed in a similar manner. In this case we take $f  = \lbrace u, c, t \rbrace$ or $\lbrace d, s, b \rbrace$ depending on which type of quarks we wish to couple $F$ (and $s$) to. In the first case, $F$ transforms as $(\mathbf{3},\mathbf{1},\mathbf{-2/3})$ (``heavy up-type quark'') and in the second one as $(\mathbf{3},\mathbf{1},\mathbf{1/3})$ (``heavy down-type quark'') under $SU(3)_c \times SU(2)_L \times U(1)_Y$. Given the similarity of the two models, in the remainder of this paper we will focus on the up-type case. We will, moreover, neglect couplings to third generation fermions\footnote{Analyzing the third generation would entail additional complications. A displaced top would be a highly complex object to reconstruct, and it is not clear how well the $b$-tagging algorithms (which are based on displacements of $B-$mesons) would perform for displaced jets. Moreover, due to the mass difference between $t$ and $b$ it is clear that the phenomenology would not be equivalent and both setups would require a different, dedicated analysis. The study of those interesting signatures is beyond the scope of this work.}.

The situation with quark-flavor bounds is similar to the one for the leptonic model. Meson-mixing constraints would stem from box diagrams (proportional to $(y_s^f)^{4}$) and penguin diagrams analogous to the $\mu \to e \gamma$ case (proportional to $(y_s^f)^{2}$), where now the photon could also be an off-shell $Z$ or a gluon. Last, but not least, one could have rare decays with invisible final states, for instance $K^+ \to \pi^+ s s$; once again these diagrams would contribute proportional to $(y_s^f)^{2}$ and we can safely neglect them\footnote{NA62 could test this rare decay up to a SM-like branching ratio of ${\cal O}(10^{-11})$, thus it is only sensitive to $y_s \gtrsim 10^{-5}$, which is well above the values required by the freeze-in mechanism.}.

Compared to the leptonic model, electroweak precision data considerations are replaced here by the running of the strong coupling, however the corresponding bounds are rather mild and in the range of a few hundred GeVs~\cite{Llorente:2018wup}. LHC searches for multi-jets plus missing energy are subdominant, as the jet requirements force them to be mostly prompt.
Finally, we note that in this setup the LHC vector-like fermion pair production does not only proceed via s-channel gluon exchange, but also via a t-channel exchange of $F$ fermions\footnote{The t-channel exchange using the freeze-in vertex $F-f-s$ is  irrelevant due to the size of $y_s^f$.}.

\vspace{2mm}

In a nutshell, the models described here are only mildly constrained from indirect effects and direct prompt searches. They constitute, hence, a favourable playground for LLP searches at the LHC.
\section{The freeze-in mechanism and cosmological considerations}
\label{sec:cosmo}

We now move to discuss some general features of the freeze-in mechanism, in particular related to the lifetime of the $F$ field, a quantity which is crucial for the collider phenomenology of our scenarios. We moreover discuss cosmological constraints on our models as well as an interesting connection that could be established between DM and baryogenesis if the decaying particles are observed at the LHC.

\subsection{Freeze-in dark matter production and parent particle lifetime}
\label{sec:FreezeIn}

In our scenarios, DM is mainly produced via the decay of the vector-like fermion $F$. Scattering processes are found to provide a subleading contribution to the total DM abundance unless the DM mass is very close to the heavy fermion one (see also \cite{Hall:2009bx,Belanger:2018ccd}\footnote{Note that, especially in the case of our models with heavy vector-like quarks, effects such as Sommerfeld enhancement could affect the annihilation-induced DM production 
rate, as well as the abundance of 
the fermion $F$. However, since in whatever follows we will stick to scenarios in which DM production through decays is kinematically wide-open, we can safely ignore such effects.}). The Boltzmann equation for the DM number density reads
\bea\label{eq:Boltz1}
\dot{n}_s + 3 H n_s = \sum_i \int \dfrac{d^3p_F}{(2\pi)^3 2E_F} \dfrac{d^3p_i}{(2\pi)^3 2E_i} \dfrac{d^3p_s}{(2\pi)^3 2E_s} (2\pi)^4 \times \nonumber \\  \delta^{(4)}(P_F-P_i-P_s)|{\cal M}_i|^2 \times [f_F (1-f_i)(1+f_s) - f_i f_s (1- f_F)] \, ,
\eea
where $f_i$ are the distribution function and $P_i = (E_i,p_i)$  are the four-momentum of the particles of type $i$. The amplitude for the processes under consideration is denoted as ${\cal M}_i$.
\\
\\
In order to proceed we will further assume the following:
\begin{itemize}
    \item The initial abundance of $s$ is zero, $n_s = 0$ which, as mentioned earlier, together with the requirement of very small couplings allows us to neglect the annihilation term on the right-hand side of equation~\eqref{eq:Boltz1}.
    \item  DM production takes place during the era of radiation domination\footnote{For freeze-in production in scenarios with a modified thermal history see \textit{e.g.} \cite{Co:2015pka}. }.
    \item The distribution of all the particles is taken to be Maxwell-Boltzmann\footnote{See \cite{Belanger:2018ccd} and \cite{Blennow:2013jba} for a discussion on the impact of the bath particle distribution functions. }.
\end{itemize}
Under these simplifying assumptions, the comoving DM number density (or yield $Y_s$) can be approximated as:
\be 
Y_s \approx \dfrac{45\,\xi\, M_{\rm Pl}}{8\pi^4\cdot 1.66} \dfrac{g_F}{m_F^2}\Gamma \int_{m_F/T_R}^{m_F/T_0} dx~x^3 \dfrac{K_1(x)}{g^s_*(m_F/x)\sqrt{g_*(m_F/x)}},
\label{eq:DMyield}
\ee
where $x=m_F/T$, $\xi=2$ since the decaying particle $F$ is not self-conjugate (otherwise $\xi=1$), $g_F$ are the internal degrees of freedom of $F$ and $\Gamma$ is the sum of all partial decay widths into DM. Moreover, $M_{\rm Pl}$ is the Planck mass and equals $1.2\times 10^{19}$ GeV, and $T_R$ is the reheating temperature of the Universe (in our context, the temperature at which DM production starts), whereas $T_0$ is the temperature today. The function $K_1(x)$ is the modified Bessel function of the second kind of degree one, and $g_*$, $g^s_*$ are the effective degrees of freedom for the energy and entropy densities, respectively. The present-day DM abundance is related with the yield through \cite{Hall:2009bx}
\be 
\Omega_s h^2\approx \dfrac{m_s Y_s}{3.6\times 10^{-9} \ {\rm GeV}}~.
\ee
In Eq.~\eqref{eq:DMyield}, the quantity most affected by the details of the underlying particle physics model is the decay width $\Gamma$.
 Consequently, by assuming that $F$ decays constitute the dominant DM production mechanism (something which we have also verified numerically), we can obtain a fairly model-independent estimate of the relation between the LLP lifetime and the LLP and DM masses by requiring that the freeze-in DM abundance meets the Planck measurement:
\begin{align} 
\label{eq:ctau}
c\tau [{\rm m} ] \approx~&4.5~\xi~g_F~\left(\dfrac{0.12}{\Omega_s h^2}\right) 
\left(\dfrac{m_s}{100~{\rm keV}}\right)\left(\dfrac{200~{\rm GeV}}{m_F}\right)^2 \\ \nonumber 
& \left(\dfrac{102}{g_*(m_F/3)}\right)^{3/2}
\left[\dfrac{\int_{m_F/T_R}^{m_F/T_0} dx~x^3 K_1(x)}{3\pi/2}\right]~,
\end{align}
where we have taken $g_*^s(m_F/x) = g_*(m_F/x)$ evaluated at $x=3$ \footnote{This is a fairly good approximation for our models, since most of the DM production occurs around the freeze-in temperature $T\approx m_F/3$.}. For the collider analysis, we will 
employ the decay length $c \tau$ and the branching fractions of $F$ instead of the Yukawa couplings $y_{s}^f$. 

From Eq.~\eqref{eq:ctau} we see that, provided that $m_F \ll T_R$ \footnote{For $T_0\ll m_F\ll T_R$, the ratio in squared brackets in Eq.\eqref{eq:ctau} will approach 1.}, obtaining the correct DM relic abundance favours rather long lifetimes for the parent particle $F$, unless a large hierarchy exists between $m_F$ and the DM mass. This observation is crucial for the LHC phenomenology of our models, since it implies that throughout most of the Planck-compatible parameter space, the $F$ particles are expected to be detector-stable. However, an alternative possibility exists that would also amount to a reduction of the parent particle lifetime: a freeze-in solution for DM masses $m_s\gg$ MeV could be obtained by decreasing the reheating temperature, such that $m_F\gtrsim T_R$. This means essentially that the DM production history is shorter, relying only on the Boltzmann tail of the parent particle. This could have important implications for baryogenesis and leptogenesis that we discuss in section~\ref{sec:baryogenesis}.

\subsection{Cosmological bounds}

Let us now briefly discuss two important cosmological constraints that our models are subject to. The first is related to the possible wash-out of small and intermediate scale structures if DM possesses a non-negligible velocity dispersion. The most stringent limits in this case are Lyman-$\alpha$ forest observations by means of which, limits ranging from $m_{\rm DM} \gtrsim 4.09$ keV \cite{Baur:2015jsy} up to $m_{\rm DM} \gtrsim 5.3$ keV \cite{Irsic:2017ixq} have been obtained for warm DM produced via conventional thermal freeze-out. An intermediate value $m_{\rm DM} \gtrsim 4.65$ keV \cite{Yeche:2017upn,Baur:2017stq} was used in \cite{Heeck:2017xbu,Boulebnane:2017fxw} in order to translate this limit to the case of freeze-in DM produced via two-body decays of a parent particle in thermal equilibrium with the plasma 
. The corresponding limit reads
\begin{equation}
m_{\rm DM} \gtrsim 12 \ {\rm keV} \left( \frac{\sum_{i} {\rm BR}_i \Delta_i^\eta}{\sum_{i} {\rm BR}_i} \right)^{1/\eta} \, ,
\end{equation}
where the sum runs over all decay channels of the type $F \to X_{\rm SM}^i + s$ that can contribute to DM production, each with a branching ratio ${\rm BR}_i$.
The parameter $\Delta_i$ is defined as $\Delta_i = 1 - m_{ X_{\rm SM}^i}^2/m_{Y}^2$ and $\eta \simeq 1.9$ \cite{Boulebnane:2017fxw}. 
Since in what follows we will consistently stick to parent particle ($F$) masses lying in the hundreds of GeV and we focus on couplings to light fermions, we have $\Delta_i \simeq 1$ for all decay channels and for all models. Moreover, since our $F$ is $\mathbf{Z}_2$-odd and the only lighter $\mathbf{Z}_2$-odd particle is the DM $s$, we also have $\sum_{i} {\rm BR}_i = 1$. This means that the quantity in parentheses is approximately unity, and an approximate lower bound 
\begin{equation}
m_s \gtrsim 12 \ {\rm keV}
\label{eq:lymanalpha}
\end{equation}
applies to our scenarios.

Additional constraints on models in which DM is produced through the decay of a heavier particle stem from the measurement of the abundances of light elements in the Universe. With the potential exception of $^7$Li, standard Big Bang Nucleosynthesis (BBN) explains these abundances with a remarkable accuracy~\cite{Iocco:2008va,Cyburt:2015mya}. If $F$ decays sufficiently late, its decay products may induce several processes that alter the predictions of BBN.

As we will see in Section~\ref{sec:collider}, in our collider analysis we will be dealing with decay lengths $c\tau$ ranging from about $1$ cm up to $10^4$ m. In a radiation-dominated Universe, and taking for simplicity $g_* \sim 100$, the latter corresponds to a lifetime of about $ 3\times 10^{-5}$ sec, \textit{i.e.} roughly a temperature of $150$ MeV. This temperature (as well as the temperature of $F$ freeze-out, for the $m_F$ values we will consider) is much higher than the one of neutrino freeze-out ($T_\nu \sim 3$ MeV) and --- even more so   --- neutron freeze-out ($T_D \sim 0.7$ MeV), implying that in everything that follows the heavy fermions decay well before the onset of BBN.

\subsection{Possibility to falsify models of baryogenesis}
\label{sec:baryogenesis}
Besides the nature of DM, the baryon asymmetry of our Universe is another long standing puzzle of modern physics and points towards BSM physics. Experimentally, the asymmetry is determined very precisely in terms of the baryon-to-photon ratio $ \eta_B^\text{obs} = \frac{n_B - n_{\bar{B}}}{n_\gamma} = \left(6.09 \pm 0.06\right) \times 10^{-10}$~\cite{Ade:2015xua}. Theoretically, it is well-established that the three Sakharov conditions~\cite{Sakharov:1967dj} have to be fulfilled: (1) baryon number (B) violation and (2) C and CP violation have to occur (3) out of equilibrium.

Only condition (1) is sufficiently realised in the SM. The chiral nature of the electroweak interactions leads to anomalies that violate baryon and lepton number (B+L) at the quantum level, leading to an infinite vacuum structure. At 
temperatures above the electroweak scale in the early Universe, transitions between different vacua (sphaleron transitions) are efficient and baryon and lepton number are violated by $\Delta(B+L)=6$ \cite{Kuzmin:1985mm}. Conditions (2) and (3), however, are not sufficiently satisfied within the SM. Quantified in terms of the Jarlskog invariant~\cite{Jarlskog:1985cw}, the amount of CP violation is not sufficient \cite{Gavela:1993ts,Huet:1994jb,Gavela:1994dt} and given the measured Higgs mass, the electroweak phase transition cannot be 
of first-order (as needed for the departure from equilibrium) within the SM itself \cite{Bochkarev:1987wf,Kajantie:1995kf}. Thus, models have to provide mechanisms involving new physics in order to overcome these limitations. 

Most popular realisations include, for instance, electroweak baryogenesis \cite{Dolgov:1991fr,Turok:1992jp,Cohen:1993nk,Rubakov:1996vz} or baryogenesis via leptogenesis \cite{Fukugita:1986hr}. While they provide new sources of CP violation and an out-of-equilibrium mechanism, they rely on efficient SM sphaleron transitions to generate a baryon asymmetry. 
In the SM, sphaleron transitions are only efficient above a certain  temperature $T^*$ --- the freeze-out temperature below which the Hubble rate is larger than the B+L violation  rate --- determined by lattice computations to be $T^* = (131.7 \pm 2.4)~\mathrm{GeV}$~\cite{DOnofrio:2014rug}. 
As a result, the $B-L$ asymmetry generated in leptogenesis scenarios has to be generated above $T^*$ such that the lepton asymmetry 
can be efficiently transmitted via sphaleron transitions into a baryon asymmetry. A reheating temperature $T_R < T^*$ would exclude such a scenario, 
falsifying leptogenesis models. In addition, if the electroweak phase transition would be of first-order (possible in certain extensions of the SM, 
see~\cite{Morrissey:2012db} for a review), the sphaleron transitions would be shut-off for temperatures lower than the transition temperature $T_{EW}$.
Electroweak baryogenesis relies on efficient sphaleron transitions which, during a first-order electroweak phase transition, 
translate the CP-asymmetry created in front of the Higgs bubble walls into a baryon asymmetry (see~\cite{Morrissey:2012db} for details) before 
being switched-off at $T < T_{EW}$. With $T_R < T_{EW}$, this mechanism too would be excluded. 

In this context our model is highly interesting, as it allows to determine $T_R$ under the assumption that $s$ makes up all of DM. In case of an observation of a long-lived particle decay that fixes the observables $c\tau$ and $m_F$, we are left with the free parameters $m_s$ and $T_R$, \textit{c.f.} Eq.~\eqref{eq:ctau}. However, it will be difficult to determine $m_s$, the mass of the DM particle. Thus, our most conservative assumption for estimating $T_R$ is to assume the lightest possible DM mass of $12~\mathrm{keV}$ as discussed around Eq.~\eqref{eq:lymanalpha}, as higher DM masses will always imply a lower reheating temperature. Even if we were to assume that our DM candidate does not account for the full relic density, our approach is conservative, as a lower contribution to the relic abundance would also imply a lower $T_R$. 

\vspace{2mm}

This yields the powerful possibility to falsify baryogenesis models that rely on efficient sphaleron transitions, 
if the observables $c\tau$ and $m_F$ point to $T_R < T^*, T_{EW}$. With the value of $T^*, T_{EW}$ depending on the specific baryogenesis model in question, 
we will consider different temperatures in our study.  We take $T^{SM}_{EW} \approx 159.6~\mathrm{GeV}$ as a maximal 
value~\cite{DOnofrio:2014rug,PhysRevD.93.025003} and $50$ GeV as an exemplary value for a super-cooled 
scenario\footnote{We note that for particular extensions of the SM, as discussed in~\cite{Konstandin:2011dr,vonHarling:2017yew,PhysRevLett.119.141301}, 
even $T_{EW} \sim 100~\mathrm{MeV}$ could in principle be possible, corresponding to extremely super-cooled scenarios. This is however far from the 
generic expectation in electroweak baryogenesis scenarios.}.
Thus, in order to interpolate between these values, we will use 
three choices of the reheating 
temperature $T_{R} = \{50, 100, 160\}~\mathrm{GeV}$ when discussing our LHC results in Section \ref{subsec:results}.

\section{Collider Analysis}
\label{sec:collider}

Let us now turn to the set of long-lived searches at the LHC that constrain the minimal models of freeze-in introduced in Section~\ref{subsec:models}. We first present each analysis, summarizing the strategy, and use the observed bounds to produce the \emph{current} exclusion limits on our models. HSCP searches~\cite{Chatrchyan:2013oca,CMS-PAS-EXO-16-036} are presented in Section \ref{subsec:HSCP} and the disappearing track analyses~\cite{Aad:2013yna,Aaboud:2017mpt,CMS:2014gxa,Sirunyan:2018ldc} in Section~\ref{subsec:DT}. For visible displaced decays of the charged parent within the detector, we use both the displaced lepton~\cite{CMS:emu8TeV,CMS:2016isf} and the displaced vertex~\cite{Aaboud:2017iio} searches, discussed in Section~\ref{subsec:DL} and~\ref{subsec:DJ}, respectively. We collect the results on current bounds in Section~\ref{subsec:results}. Finally, we attempt a naive extrapolation to the High Luminosity LHC (HL-LHC), presented in Section~\ref{subsec:HL-LHC}. The summary of these searches together with their total integrated luminosities can be found in Table~\ref{tab:LHCsearches}.

\begin{table}
\centering
\begin{tabular}{|c|l|l|l|l|}
\hline
\multicolumn{1}{|l|}{Search type} & Collaboration          & $\sqrt{s}$ [TeV] & $\cal L$ {[}fb$^{-1}${]} & Ref.    \\ \hline
\multirow{2}{*}{HSCP} & \multirow{2}{*}{CMS}   & 8   & 18.8          & \cite{Chatrchyan:2013oca} \\
                      &                        & 13  & 12.9          & \cite{CMS-PAS-EXO-16-036} \\ \hline
\multirow{2}{*}{DT}   & ATLAS & 13   & 36.1          & \cite{Aaboud:2017mpt}       \\
                                 & CMS    & 13   & 38.4          & \cite{Sirunyan:2018ldc}       \\ \hline
  \multirow{2}{*}{DL}   & \multirow{2}{*}{CMS}   & 8   & 19.7          & \cite{CMS:emu8TeV}        \\
                      &                        & 13  & 2.6           & \cite{CMS:2016isf}        \\ \hline
\multicolumn{1}{|l|}{DV+MET}      & ATLAS      & 13  & 32.8          & \cite{Aaboud:2017iio}     \\ \hline
\end{tabular}
\caption{Summary of all LHC searches recasted in this work: Heavy Stable Charged Particles (HSCP), Disappearing Tracks (DT), Displaced Leptons (DL) and Displaced vertices plus MET (DV+MET). We indicate the center-of-mass energy of the run and the integrated luminosity of each dataset. See main text for details.}
\label{tab:LHCsearches}
\end{table}

\subsection{Searches for Heavy Stable Charged Particles (HSCP)}
\label{subsec:HSCP}

The FIMP models discussed in Section~\ref{subsec:models} contain a charged particle $F$ which, for some of the cosmologically viable scenarios discussed in Section~\ref{sec:cosmo}, can be stable at collider scales and decay outside the detector. In this case, after being pair produced $F$ can be considered as a heavy stable charged particle (HSCP) for sufficiently large lifetimes. We must stress, however, that the HSCP signature strongly depends on the nature of $F$: if it is colour-neutral, as in the leptonic model, its interaction with the inner tracker will appear as an anomalous ionizing track. On the other hand, if $F$ is a colour triplet (vector-like hadronic model), it is expected to hadronize into neutral and/or charged hadrons ({\it R-hadrons}). 
The fraction of charged hadrons produced depends on the hadronization model considered and will affect the HSCP limits. Furthermore, as it interacts with the detector (mostly in the calorimeter), the heavy hadrons may flip charge. For simplicity, from here on we will only consider the constraints obtained for the cloud model hadronization, as described in Ref.~\cite{Chatrchyan:2013oca} and references therein.

When a massive charged particle, such as the lepton-like $F$ or a charged $R$-hadron, has a lifetime $c \tau \gtrsim {\cal O} (m)$,
a sizable fraction of the produced particles will decay only after crossing the tracker and/or the muon chamber.
Due to its large mass, the HSCP typically travels with  a velocity $\beta = v /c < 1$. Hence, as it traverses the detector, the charged particle produces an ionizing track with a higher ionization energy loss when compared to SM particles. Furthermore, if it decays outside the detector, the time-of-flight (TOF) measured using the hits in the muon chamber will be larger than that for relativistic muons\footnote{For recent ideas on exploiting the timing difference due to BSM particles, see~\cite{Liu:2018wte,Cerri:2018rkm}.}.
These two features can be used to efficiently distinguish non-SM HSCPs from other SM particles, resulting in searches with very low background.

HSCP searches using the signatures described above have been performed both by ATLAS and CMS using data from 7, 8 and 13 TeV center-of-mass energies~\cite{Aad:2012pra,Chatrchyan:2012sp,ATLAS:2014fka,Chatrchyan:2013oca,Aaboud:2016dgf,CMS:2016ybj}. 
These searches have presented their results within supersymmetric models containing long-lived squarks, gluinos or sleptons. Since the CMS constraints are similar to the ones obtained by ATLAS, and the former provide more details for recasting, in the following we will only consider the limits obtained by the CMS searches from  Refs.~\cite{Chatrchyan:2013oca} and \cite{CMS-PAS-EXO-16-036}.
Furthermore, in Ref.~\cite{Khachatryan:2015lla}, detailed trigger and selection efficiencies have been provided for recasting the 8~TeV CMS search for HSCPs. These efficiencies will be very useful when re-interpreting these searches for the FIMP scenario, as discussed below.

The CMS HSCP searches observed no excess over the expected SM background, resulting in upper limits for the production cross-section of HSCPs. In particular, limits were obtained for the direct pair production of long-lived stops, staus and heavy lepton-like fermions, assuming that these decay outside the tracker ({\it tracker-only analysis}) or outside the muon chamber ({\it tracker+TOF analysis}). Although the tracker+TOF limit is usually more constraining, 
it requires that the HSCP decays outside the detector and is only relevant for $c \tau \gtrsim 10$~m.
Since the long-lived stau and stop scenarios considered by CMS 
cover the mass window relevant for the FIMP models presented in Section~\ref{subsec:models} and
the signal efficiencies are weakly dependent on the HSCP spin~\cite{Chatrchyan:2013oca,CMS-PAS-EXO-16-036}, we can use the stau and stop limits obtained by CMS to constrain the corresponding FIMP models.
However, before we can apply the cross-section limits from the HSCP searches to the FIMP scenarios, we must account for the finite lifetime of the vector-like fermion $F$, which can vary over a wide range of values.
Since the CMS limits apply to the fraction of charged particles which decay outside the CMS tracker or muon chamber,
we must take into account the suppression of the signal due to $F$ decays taking place before the tracker (for the  tracker-only analysis) or the muon chamber (for the tracker+TOF analysis). The inclusion of this effect will be discussed next.

If $F$ has a lifetime $c\tau \sim 0.01$~m, a large fraction of the $F$ decays will happen within
the tracker, thus suppressing the HSCP signal\footnote{For simplicity we assume zero efficiency for reconstructing or selecting the HSCP track if it stops within the tracker (for the tracker-only analysis) or within the detector (for the tracker+TOF analysis).}.
For such small lifetimes we must rescale the $F$ production cross-section by the effective fraction of $F$ particles which decay outside the tracker (for the tracker-only analysis) or outside the CMS detector (for the tracker+TOF analysis). This effective fraction ($f_{LLP}$) can be easily computed using hadron-level events and the trigger and selection efficiencies from Ref.~\cite{Khachatryan:2015lla} (see Appendix~\ref{app:HSCP} for details).
Once $f_{LLP}(L,\tau)$ is known, the total cross-section for pair production of $F$s can be rescaled as:
\be
\sigma_{eff} = \sigma \times f_{LLP}(L,\tau),
\ee
where $\tau$ is the $F$ lifetime, $\sigma$ its total pair production cross-section and $L$ the relevant detector size. For the tracker-only analysis we take $L = 3$~m while for the tracker+TOF analysis we take $L=11$~m. The production cross section is computed at tree level using {\sc MadGraph5\_aMC@NLO}. The effective cross-section, $\sigma_{eff}$, can then be directly compared to the cross-section upper limits obtained by CMS for the direct production of staus (for the leptonic FIMP scenario) and stops (for the hadronic scenario) from Refs.~\cite{Chatrchyan:2013oca} and \cite{CMS-PAS-EXO-16-036}.

\begin{figure}[htb]
\centering
\includegraphics[width=0.49\textwidth]{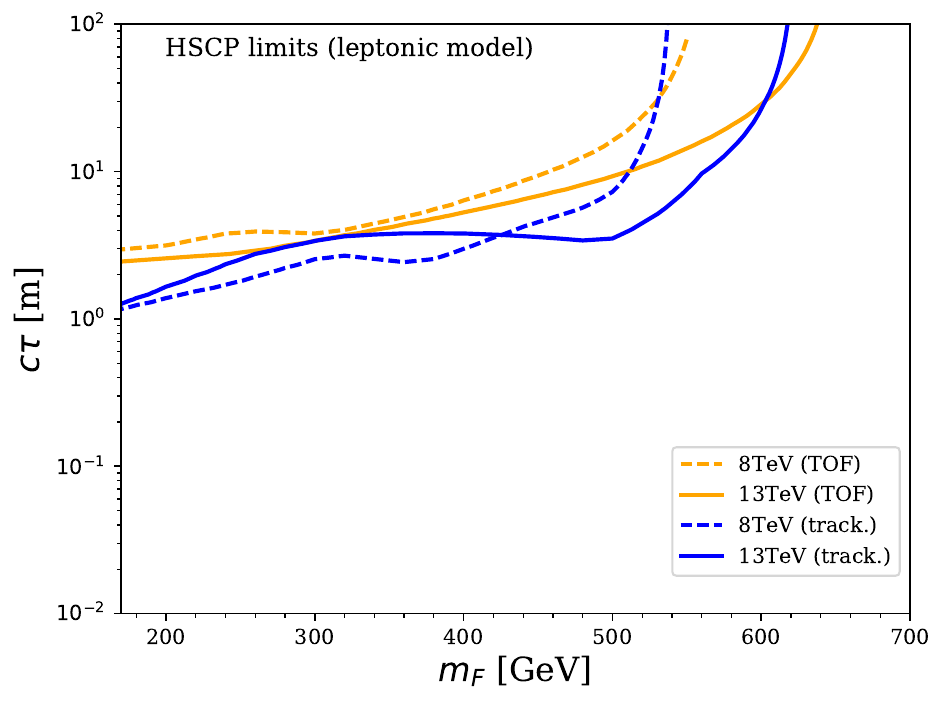}
\includegraphics[width=0.49\textwidth]{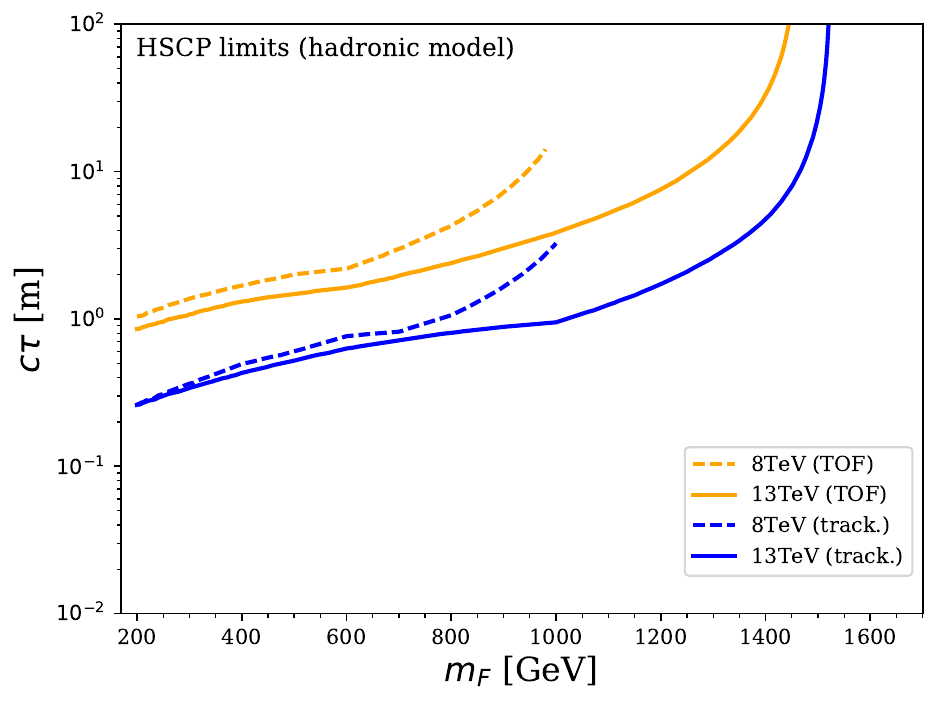}
\caption{Region in the $c \tau$ vs $m_F$ plane excluded at 95\% C.L. by the CMS 8 TeV (18.8 fb$^{-1}$) and 13 TeV (12.9 fb$^{-1}$)  HSCP searches.
Left: exclusion for the lepton-like scenario.
Right: exclusion for the quark-like scenario.}
\label{fig:HSCP_curves}
\end{figure}

In Figure~\ref{fig:HSCP_curves} we show in the $m_{F}$ vs $c \tau$ plane the exclusion curves for both the tracker-only and the tracker+TOF analyses when applied to the lepton-like and quark-like FIMP scenarios.
The leptonic model is excluded up to $m_{F} \simeq 650$~GeV for $c \tau \gtrsim 100$~m, with the exclusion dropping quickly as the lifetime is reduced\footnote{The excluded values of $m_F$ for the lepton-like FIMP scenario are higher than the CMS limits for pair production of staus, due to the larger cross-section for $F$ pair production.}.
Although the 13~TeV CMS analysis is the most constraining one for large $c \tau$, the 8~TeV analysis, which has a higher integrated luminosity, is slightly more constraining for small $c \tau$.
As expected, the tracker+TOF analysis is more constraining
for large lifetimes ($c \tau \gtrsim 50$~m), for which most of the produced $F$ particles decay outside the detector. For smaller lifetimes, the tracker-only analysis results in stronger limits, since in this case the HSCPs are only required to decay after crossing the tracker.
The corresponding limits for the hadronic model are shown in the right panel of Figure~\ref{fig:HSCP_curves}.
In this case the exclusion extends up to $m_F \simeq 1.5$~TeV for $c \tau > 100$~m. However, unlike the lepton-like case, the strongest constraints are always given by the 13~TeV tracker-only analysis.
This is caused by the fact that heavy hadrons ({\it R-hadrons}) formed with the long-lived coloured $F$ can flip charge as they traverse the detector, resulting in a neutral {\it R-hadron}. Thus events which pass the tracker-only selection may fail the tracker+TOF selection even for $c \tau \gg 100$~m, due to the charge flipping effect.
Finally,  we point out that the sharp drops seen in the 8~TeV  curves at high $m_F$ are simply due to the limited interval of masses considered, as the analysis did not provide limits for $m_F \gtrsim 550$~GeV. Similarly the 13 TeV study did not quote limits for $m_F \lesssim 160$~GeV.

\subsection{Disappearing tracks (DT)}
\label{subsec:DT}

To address the sensitivity gap between the HSCP searches targeting long lifetimes and  the displaced lepton search (to be described in section~\ref{subsec:DL}) targeting short lifetimes in the leptonic model, we now investigate the disappearing track search.
 We consider the 13 TeV analyses by ATLAS~\cite{Aaboud:2017mpt} and CMS~\cite{Sirunyan:2018ldc}, performed with total integrated luminosities of  (36.1) fb$^{-1}$ in the ATLAS case and 138.4 fb$^{-1}$ in the CMS case. These studies considered a final state with one disappearing track and at least one jet of high transverse momentum ($p_T \gtrsim 100~\rm{GeV}$) to trigger on the event.
Additional selection cuts impose a minimum angle in the transverse plane $\Delta \phi > 0.5$ between the hardest jet and the transverse missing energy vector, and ATLAS also imposes a lepton veto. The key object to identify is an isolated track reconstructed in the pixel and strip detectors having missing hits in the outer tracker (CMS), or a track with only pixel hits (ATLAS). Due to the addition of an insertable B-layer~\cite{CERN-LHCC-2012-009,Abbott:2018ikt}
 in the shutdown between Run 1 and Run 2, ATLAS can reconstruct tracks as short as 12 cm, while for the CMS tracker this distance is about 25-30 cm, depending on the direction of the track. Hence, the ATLAS search can reach lower lifetimes than the CMS study, whereas CMS sets stronger constraints at larger $c \tau$.
As a result, ATLAS can exclude down to 2-3 times lower lifetimes for the same mass in the benchmark AMSB wino model with mass up to 500 GeV, but the two searches have a comparable sensitivity for $c \tau \sim 50$ cm, and CMS has a better coverage for $c \tau \gtrsim 1 m$.

One further complication of applying the ATLAS analysis is its reconstructed lepton veto.  Since the leptons produced in $F \rightarrow l s$ are in principle also displaced and do not reconstruct back to the primary vertex, it is possible that most events will survive this veto.  Moreover, the presence of this lepton also affects the kinematics of the event by altering the missing energy distribution and adding energy deposits in the calorimeter. Hence, in order to accurately use the ATLAS search it would not be enough to estimate the probability of not reconstructing the lepton, but also the effect in the relevant distributions. A more detailed study of these effects is certainly warranted but cannot be made with only the Monte Carlo tools and efficiencies available to the theorist. In the present work, we will simply assume that the lepton always fails reconstruction, which will lead to the strongest possible bounds from this search. We also stress that the ATLAS search also implements a strict muon veto by requiring no activity in the muon system, which would lead to events where $F \rightarrow \mu s$ failing the event selection with a high probability.  The ATLAS exclusions therefore also correspond to the case where $Br(F\rightarrow e s) = 1.0$.

The CMS analysis~\cite{Sirunyan:2018ldc} does not impose a lepton veto, but instead requires that there are no substantial energy deposits in the calorimeter within a cone of radius $R = 0.5$ around the selected high-$p_T$ track. We note that for the case of $F$ decays into $e s$ such requirement also likely fails for some events.  However, we again neglect the effect of the isolation requirement as the lepton is not prompt. We do not assume particular branching ratios as there is no specific muon veto.  However events with muons are more likely to pass the isolation requirement due to smaller energy deposits in the calorimeter, thus making the CMS search complementary to the ATLAS one.

To recast the limits from both ATLAS and CMS, we make use of the generator-level efficiency tables provided by the respective experimental analyses~\cite{Aaboud:2017mpt,Sirunyan:2018ldc}.  They contain in particular, the 2D distribution of event selection efficiency in resonance mass and lifetime.  The total number of events is simply calculated as a product of the production cross section times efficiency and luminosity ($\mathcal{N} = \sigma_\mathrm{pp\rightarrow F \bar F} \times \varepsilon(m,\tau) \times \mathcal{L}$). The exclusions in parameter space are calculated using the method described in appendix~\ref{app:stat}, and shown in 
Figure~\ref{fig:DT_curve}. From the Figure we see that our intuition was indeed correct, and that ATLAS is stronger for $c \tau \sim$ 20 cm excluding $m_F$ up to 275 GeV, while CMS provides a stronger exclusion for $c \tau \sim 1$ m, excluding $m_F$ below about 335 GeV. However we already know that the HSCP search is very constraining $c \tau$ above a few meters, thus stressing the limitations of the DT search for the leptonic model. We will further discuss these results in section~\ref{subsec:results}.

\begin{figure}[htb]
\centering
\includegraphics[width=0.80\textwidth]{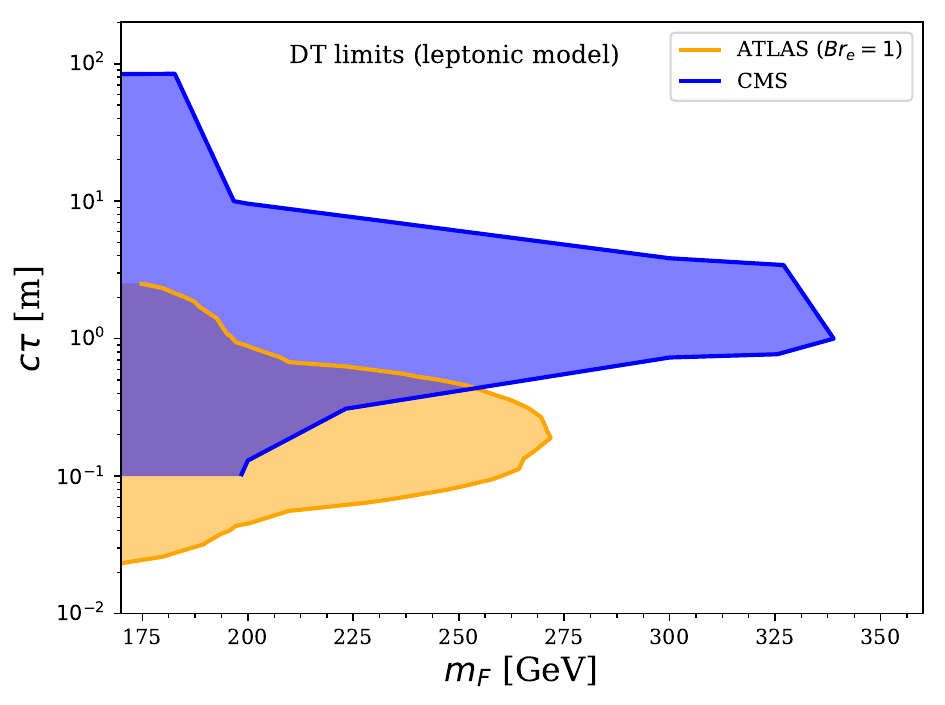}
\caption{Region in the $c \tau$ vs $m_F$ plane excluded at 95\% C.L. by disappearing tracks searches for the leptonic model. We employed the ATLAS 13 TeV (36.1 fb$^{-1}$) and CMS 8 TeV (19.5 fb$^{-1}$) studies.}
\label{fig:DT_curve}
\end{figure}

\subsection{Displaced lepton searches (DL)}
\label{subsec:DL}

For the leptonic model from Section~\ref{section:leptonmodel}, if $F$ can decay both to electrons and muons, then the CMS searches for events with oppositely charged, displaced electrons and muons, performed at both 8 TeV~\cite{CMS:emu8TeV} (with $19.7$ fb$^{-1}$) and 13 TeV~\cite{CMS:2016isf} (with $2.6$ fb$^{-1}$) could provide relevant bounds on the model.
These searches are maximally sensitive in the regime $y^e_s \simeq y^{\mu}_s$ (which yields similar branching fractions of $F$ to electrons and muons)\footnote{An attempt to extend this search to the case of two muons or two electrons has been performed in~\cite{Evans:2016zau}.}. 

The discriminating kinematical variable in both analyses is the lepton transverse impact parameter $d^{\ell}_0$, defined as the closest distance between the beam axis and the lepton track in the transverse plane
\begin{equation}
d^{\ell}_0 = \frac{\left| p^{\ell}_x L_y - p^{\ell}_y L_x  \right|}{p^{\ell}_T} \, ,
\end{equation}
where $L_{x,y}$ are the distances in $x,y$ travelled by the LLP before decaying, $p^{\ell}_T$ is the 
transverse momentum of the lepton and $p^{\ell}_{x,y}$ the $x$, $y$ components 
of the lepton 3-momentum. 

The CMS displaced-$e\mu$ search selects events with exactly one electron and one muon, both satisfying isolation criteria\footnote{A lepton is considered ``isolated" when the sum of the $p_T$ of \emph{all other} particles (normalized to its own $p_T$) within a cone of radius $R$ is below a certain $\epsilon$ value. For electrons in the barrel, electrons in the endcap and muons $(R,\epsilon) = (0.3, 3.5 \%), (0.3, 6.5 \%)$ and $(0.4, 15 \%)$ respectively.}, with $\left|\eta_{\ell} \right| < 2.4$ and $p^{\ell}_T > 42$ ($40$) GeV for electrons (muons), together with the requirement $\Delta R_{e\mu} > 0.5$. In addition, the LLP decay is required to occur within $L_z < 300$ mm, $\sqrt{L_x^2 + L_y^2} < 40$ mm, otherwise the tracking selection fails.
The 8 TeV CMS analysis~\cite{CMS:emu8TeV} then defines three non-overlapping signal regions (SR): 

\begin{itemize}
 
 \item SR III: Both $d^e_0$ and $d^{\mu}_0 \in [1, 20]$ mm. 
 
 \item SR II: Both $d^e_0$ and $d^{\mu}_0 > 0.5$ mm but one or both leptons fail SR III.
 
 \item SR I: Both $d^e_0$ and $d^{\mu}_0 > 0.2$ mm but one or both leptons fail SR III and SR II.
 
\end{itemize}

\noindent and publicly provides identification efficiencies  
for electrons and muons as a function of $p^{\ell}_T$ and $d^{\ell}_0$. 
The CMS 13 TeV search~\cite{CMS:2016isf} extends SR III to $d^{e,\mu}_0 \in [1, 100]$ mm, but does not publicly provide any lepton 
efficiencies\footnote{For an attempt to reproduce the results of  
the displaced-$e\mu$ 13 TeV CMS analysis~\cite{CMS:2016isf} by using the public CMS 8 TeV efficiencies from~\cite{CMS:emu8TeV}, see Cottin et al.~in ref.~\cite{Brooijmans:2018xbu}.}. 
We thus extrapolate here the public CMS 8 TeV efficiencies to the 
13 TeV analysis and do not consider the region 
$d^{e,\mu}_0 \in [20, 100]$ mm. This provides a conservative reach of the 
13 TeV displaced-$e\mu$ search. 
We use {\sc MadGraph5\_aMC@NLO}~\cite{Alwall:2014hca} to simulate $F$ pair-production from Drell-Yan, including a flat NLO $\kappa$-factor for Drell-Yan production, $\kappa \sim 1.2$~\cite{Beenakker:1999xh}.
We consider that the vector-like leptons $F$ decay to both electrons and muons, and study both the case of equal branching fractions and a $90\% - 10\%$ branching fraction scenario with either electron or muon as the dominant channel.  

\begin{figure}[htb]
\centering
\includegraphics[width=0.80\textwidth]{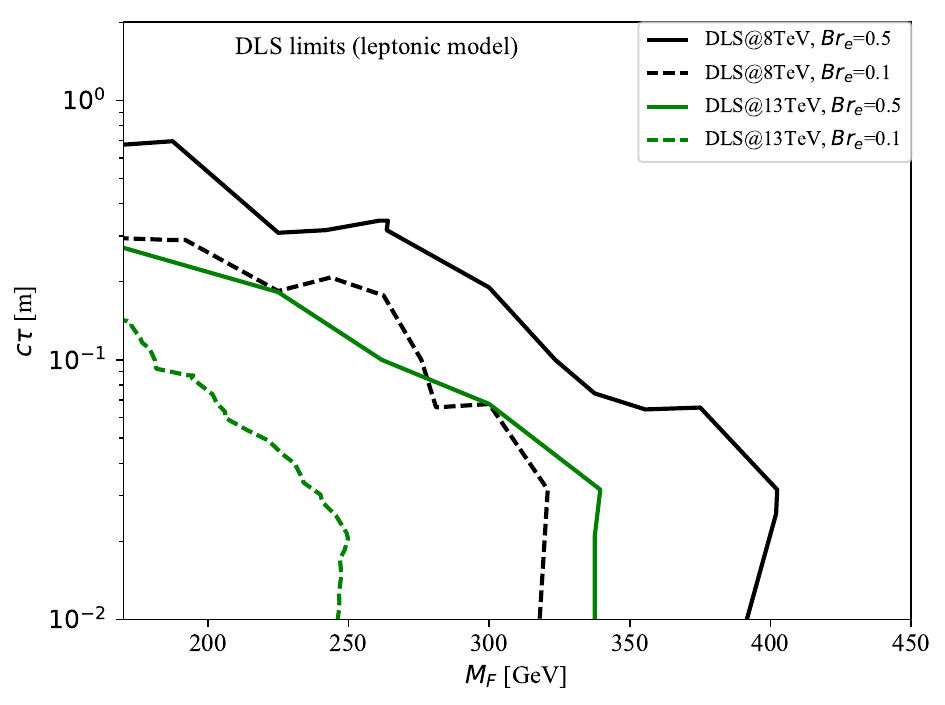}
\caption{95 \% C.L exclusions in the $c \tau$ vs $m_F$ plane due to the 8 TeV (19.7 fb$^{-1}$, black) and 13 TeV (2.6 fb$^{-1}$, green) CMS Displaced Lepton searches. The solid curve indicates the case where  $Br(F\rightarrow e s) \equiv Br_e = 0.5$ and $Br(F\rightarrow \mu s) \equiv Br_{\mu} = 0.5$, while the dashed line denotes the case where $Br_e=0.1$ and $Br_{\mu}=0.9$.}
\label{fig:DSL_curve}
\end{figure}

For a given value of $m_F$ and $c \tau$, we generate 200k MC events and apply the CMS event selection. We then compute the impact parameter of both leptons in each event and obtain the number of expected signal events in SR I, II, III, $s_i (c \tau, m_{F})$, for both 8 and 13 TeV. We subsequently compute the 95 \% C.L. exclusion limits on the leptonic model as described in appendix~\ref{app:stat}. These limits are shown in Figure~\ref{fig:DSL_curve}. 

For the 8 TeV CMS search, which currently provides the strongest limits,  
the number of expected background events in SR I, II, III are  $18.0\pm0.5\pm3.8$, $1.01 \pm 0.06 \pm 0.30$ and $0.051 \pm 0.015 \pm 0.010$ respectively (we quote first the statistical error and then the systematic error, which are then added in quadrature), with the dominant background in all three regions corresponding to heavy flavour (HF), obtained through a data driven estimate~\cite{CMS:emu8TeV}. 
For the 13 TeV analysis, the low statistics of the search only allows to set a 68 \% C.L. upper bound on the HF contribution to the background, translating into an upper bound on the total predicted background in SR I, II, III, corresponding respectively to $ < 3.2$, $< 0.50$ and $< 0.020$ events. In this case, we treat the uncertainty in the HF background as the error of our background prediction (this is also relevant for the extrapolation to HL-LHC performed in Section~\ref{subsec:HL-LHC}). 

From the figure we see that these searches have a good coverage for short lifetimes, and thus they probe complementary regions of the freeze-in parameter space than the HSCP and DT searches described before. Note that since this study requests one electron and one muon the actual coverage depends on the specific branching fractions, and should vanish if one of the two decay modes is closed. While this is an extreme possibility, it would certainly be quite interesting to carry out experimental searches dropping this requirement (see~\cite{Evans:2016zau} for an analysis along those lines).

\subsection{Displaced Vertices plus MET (DV+MET)}
\label{subsec:DJ}

As already discussed in Section~\ref{subsec:HSCP}, in the quark-like FIMP case, where $F$ is a meta-stable colour triplet, it will hadronize into neutral and/or charged hadrons ({\it R-hadrons}) before decaying. For lifetimes $c \tau \lesssim 10$~m, sensitivity to this case could be assessed through searches for displaced vertices.  Since in the quark-like scenario $F$ decays to a quark and DM, the signature corresponds to displaced jets in association with missing energy.
In Ref.~\cite{Aaboud:2017iio} ATLAS has performed a search for multi-track displaced vertices in association with large missing transverse momenta at 13~TeV and with 32.8~fb$^{-1}$ data.  The search selects events with reconstructed  $E_T^{miss} > 250$~GeV, and at least one displaced vertex with five or more tracks and a visible invariant mass greater than 10 GeV.  

Interpreting this analysis requires a highly realistic simulation of the detector response and event reconstruction, which can only be achieved reliably by the internal experimental software.  As an alternative, we use a prescription based on parametrized efficiencies provided by the ATLAS Collaboration as a function of vertex radial distance, number of tracks and mass~\cite{Aaboud:2017iio} and apply these efficiencies to the truth level signal MC events. In this selection, the truth level missing energy is required to be $E_T^{miss} > 200$~GeV.
Furthermore, events should have either one jet with $p_{T}>70$~GeV or two jets with $p_{T}>25$~GeV. In addition, events must have at least one displaced vertex consistent with the following requirements:
\begin{itemize}
\item{transverse distance between the impact parameter and the decay position $d_{xy} >4$ mm.}
\item{the decay position must lie within the fiducial region, $r_{DV}<300$~mm and $|z_{DV}|<300$~mm, where $r~(z)$ are the transverse (longitudinal) position of the DV.}
\item{the number of selected decay products must be at least 5, where selected decay products are charged and stable, with $p_{T}>1$ GeV and $|d_{0}|>2$ mm.}
\item{the invariant mass of the truth vertex must be larger than 10 GeV, and is constructed assuming all decay products have the mass of the pion.}
\end{itemize}
The above procedure was then validated on the signal scenario used by ATLAS for interpretation, which is a split SUSY simplified model with a long-lived gluino that hadronizes forming an $R$-hadron before decaying, i.e., $\tilde{g}\rightarrow q\bar{q}\tilde{\chi}^{0}_{1}$\footnote{This validation study was done earlier for the Les Houches PhysTeV 2017 proceedings~\cite{Brooijmans:2018xbu}.}. 

\begin{figure}[htb]
\centering
\includegraphics[width=0.80\textwidth]{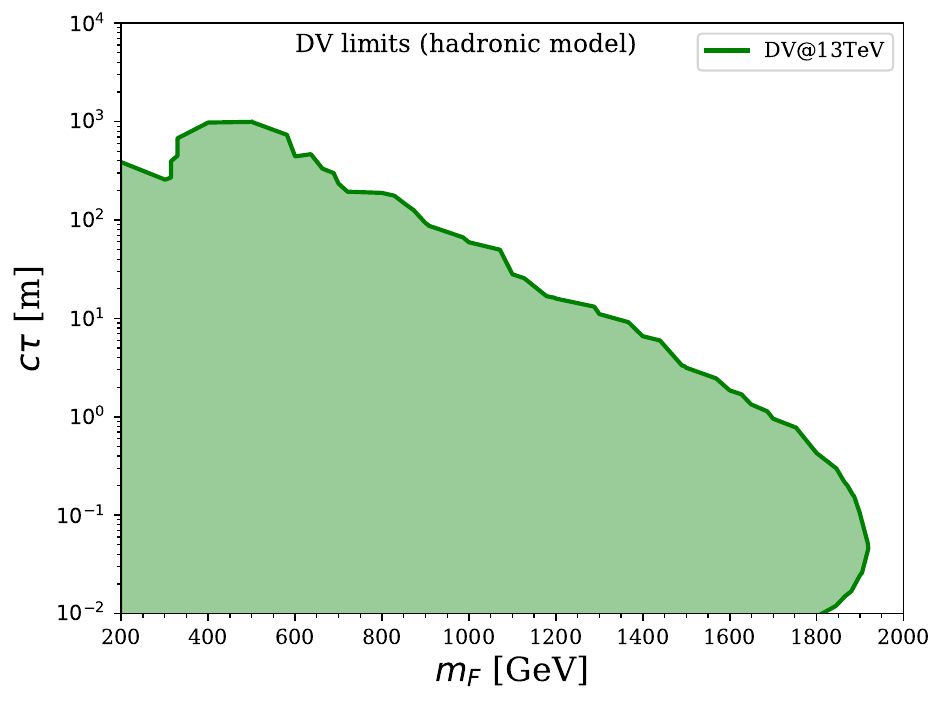}
\caption{Region in the $c \tau$ vs $m_F$ plane excluded at 95\% C.L. by the ATLAS 13 TeV displaced vertex plus MET search~\cite{Aaboud:2017iio}.}
\label{fig:DV_curve}
\end{figure}

The ATLAS search observed no excess over the expected background, resulting in
the 95\% C.L. limit of $N_{signal} < 3$ events, after the event selection described above has been applied. In order to apply this constraint to the quark-like FIMP scenario, we use  {\sc MadGraph5\_aMC@NLO}~\cite{Alwall:2014hca} and Pythia 8~\cite{Sjostrand:2014zea} to simulate the production and decay of $F$.
In order to hadronize $F$ we assume the Pythia 8 hadronization model for long-lived stops, which should be a good approximation, since the hadronization is largely spin independent.
For a given value of $m_F$ and $c \tau$, we generate 50k MC events and apply the ATLAS event selection and efficiencies in order to obtain $N_{signal}$.
In Figure~\ref{fig:DV_curve} we show the excluded region in the  $c \tau$ vs. $m_F$ plane for the quark-like scenario.
As expected, the ATLAS analysis is mostly sensitive to decays
which take place inside the tracker ($c \tau \sim$~few cm).
In particular, for $c \tau \simeq 3$~cm, the exclusion extends up to
$m_F = 1.9$~TeV. More surprisingly is the sensitivity of the ATLAS search  for lifetimes as large as $c \tau \sim 100$~m.
However, since the distribution for the decay position is a decreasing exponential, even for such large lifetimes there is still a significant fraction of {\it R-hadrons} which decay within the tracker.

At this point it is worth noting that we compared these limits with the ones obtained from prompt multi-jet + missing energy searches corresponding to the above topology. This was done by recasting the CMS 13 TeV SUSY multi-jet + missing energy search with $35.9$ fb$^{-1}$ luminosity \cite{Sirunyan:2017cwe} within the PAD (public analysis database) framework of \texttt{MadAnalaysis5} \cite{Dumont:2014tja}. The recasted analysis can be found at the link \cite{recast:prompt}. 
We observed that the prompt searches are significantly weaker than the displaced vertex + MET search, thus we ignore the former in the following.

\subsection{Results}
\label{subsec:results}

\begin{figure}[htb]
\centering
\includegraphics[width=0.99\textwidth]{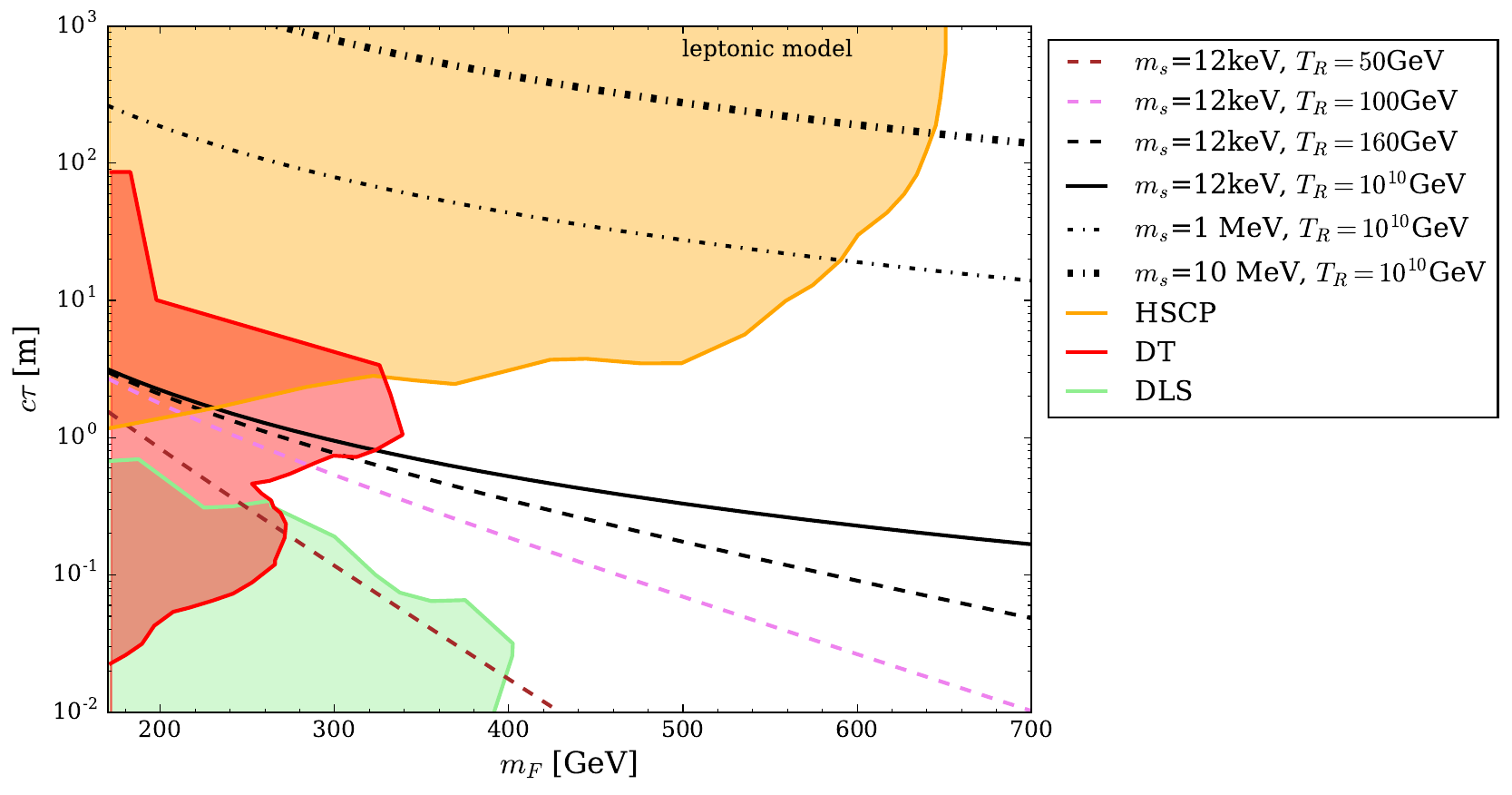}
\includegraphics[width=0.99\textwidth]{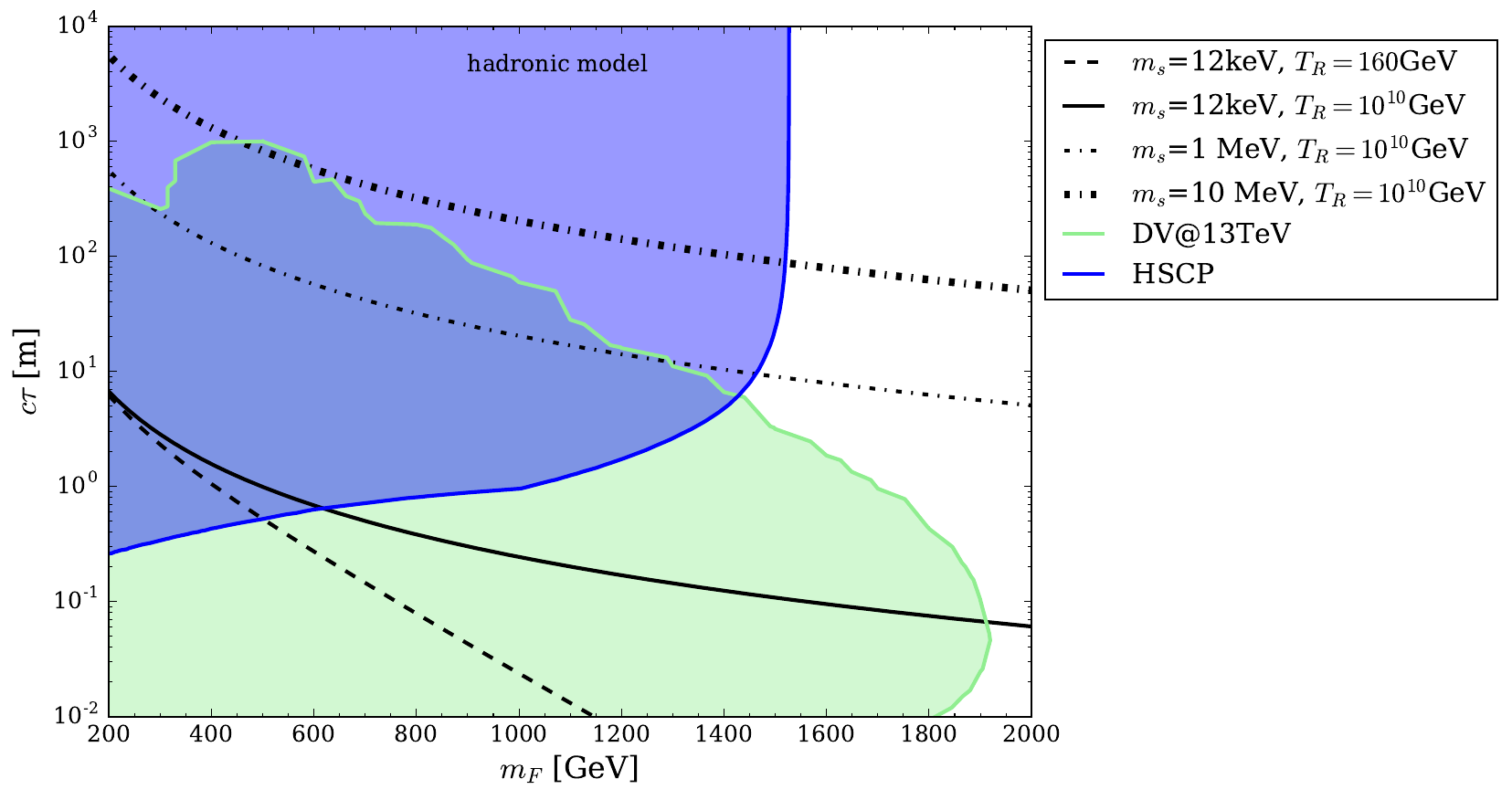}
\caption{Summary of the LHC constraints for the lepton-like (upper panel) and the quark-like (lower panel) FIMP scenarios. The lines correspond to contours of $\Omega_s h^2=0.12$ for the values of $m_s$ and $T_R$ given in the legend.}
\label{fig:summary}
\end{figure}

After having described in detail the relevant experimental searches, we collect all results for the leptonic model and hadronic model, which are displayed in upper and lower panels in Figure~\ref{fig:summary}. 
As most of these studies (HSCP, DT, DL, etc.) have several exclusion limits, due to data taken at LHC 8/13 TeV by both ATLAS and CMS, as well as to different search strategies being pursued, we present one single 95 \% C.L exclusion curve for each different type of analysis. This exclusion curve is the outer boundary (``envelope'') of the various exclusion regions in the $c\tau - m_F$ plane (in other words, for a fixed mass we will consider the largest exclusion interval for $c \tau$). Together with these exclusion curves we also 
show in Figure~\ref{fig:summary} several lines corresponding to 
the correct DM relic density (recall the discussion in Section~\ref{sec:FreezeIn}) 
for various choices of DM mass $m_s$ and reheating temperature $T_R$.

We first consider a reheating temperature much higher than all other scales in the model and choose, for concreteness, $T_R=10^{10}$ GeV (although we stress that the specific value chosen has no impact on the results). 
We 
pick three different DM masses, namely $m_s=12~{\rm keV},1~\rm{MeV}$ and $10~\rm{MeV}$ (solid, dashed and dot-dashed), where the lowest value 
is motivated by
 the bound set by Lyman-$\alpha$ observations, Eq.~\eqref{eq:lymanalpha}. For $m_s=12~{\rm keV}$, the values of 
 $T_R = 50,\,100,\,160$ GeV are chosen in order to assess the prospects of testing electroweak baryo- and leptogenesis, as discussed in Section~\ref{sec:baryogenesis}. The simultaneous determination of $c \tau$ and $m_{F}$ would then allow to infer the maximum possible reheating temperature $T_R$ by assuming the lightest possible DM mass of $m_{s} = 12$ keV. 
 
Within the leptonic model and for $T_R=10^{10}$ GeV, $F$ masses as high as 600-650 GeV can be probed for proper lifetimes of $c \tau$ larger than 20 meters (or conversely DM masses as low as a few MeV). For the smallest possible DM mass of 12 keV, the $F$ lifetime ranges between a few meters and tens of centimeters, and the excluded values of $m_F$ correspond to 300-350 GeV. For the case of a low $T_R$ the lifetime decreases and the bound on $m_F$ is reduced down to about 250 GeV for $c \tau \sim 6$ cm, thus forming a ``wedge" in the parameter space\footnote{This is an interesting region of parameter space as it can happen naturally in electroweak models of dark matter (including SUSY "pure" scenarios) and thus we expect the HL-LHC upgrades will specifically target this difficult region and significantly expand their coverage, using for instances ideas like the one developed in~\cite{Fukuda:2017jmk}.}. The DT analysis is not able to properly cover the case where most of the signal events leave only a few hits in the inner detector. Further lowering $T_R$ leads into the region that can be well covered by the DL search, and thus the mass reach goes up to 400 GeV for $c \tau \sim 1$ cm.

We now turn to the small reheating temperatures that indicate the limit below which one can probe specific baryogenesis models. 
While supercooled scenarios ($T_R < 50$ GeV) cannot be falsified with our analysis since the corresponding parameter space is almost already probed (except for a tiny region around $m_F \sim $ 400 GeV), there is still parameter space left  below 
$T_R=160$ GeV, in which an observation of a signal at the LHC could bring baryogenesis models within the freeze-in set-up in tension. Thus, future LHC results that will explore this region will be of high interest. However, we note that in case of a positive signal, the accuracy on the extraction of the reheating temperature would strongly depend on the actual value of $c \tau$. 

The worst case scenario corresponds to the upper possible value of $c \tau = 50$~cm. The corresponding masses for $T_R=50, 160, 10^{10}$ GeV are 225, 325 and 380 GeV. Thus measuring $m_F=400$ GeV with a 5\% accuracy and $c \tau$ with infinite precision would indicate $T_R \gtrsim$ 200 GeV, an accurate determination of the high-scale $T_R$ value, however, will not be possible. In contrast, with a positive signal at lower masses and/or lower lifetimes $T_R$ can be determined with good accuracy. For instance, measuring $m_F=300$ GeV with a 5 \% precision and once again the lifetime with infinite precision, $c \tau=50$, cm would allow a determination of $T_R=85\pm 10$ GeV, while having $m_F = 500$ GeV and $c \tau = 1$ cm would yield $T_R \in [58-68]$ GeV.

While discriminating between different values of higher reheating temperatures might be complicated as we just mentioned, we see that for the lower values in the 50-160 GeV interval this is an achievable task even if $m_F$ and $c \tau$ are only roughly determined.  For instance, in the displaced lepton search this could be achieved through the combined information on the total number of signal events and their $d^{\ell}_{0}$ distribution, while for the displaced vertex plus MET search this could be achieved by combining the total number of signal events and the vertex displacement information. Studying the expected accuracy of the extraction of these parameters is beyond the scope of this work\footnote{See e.g Banerjee et al in ~\cite{Brooijmans:2018xbu} for similar work on estimating the lifetime for displaced leptonic vertices.}.

For the hadronic model, we have used the same three benchmarks for $T_R=10^{10}$ GeV temperatures, and for the leptonic case we only show the case of $T_R=160$ GeV, 
since the DV+MET search from ATLAS already excludes the $m_s=12$ keV scenarios with low reheating temperatures. This clearly demonstrates that within this model, 
electroweak baryogenesis cannot be excluded anymore for the most conservative assumption of $m_s=12$ keV based on our considerations on the reheating temperature. 
For $T=10^{10}$ GeV they force $m_F$ above 1.5 TeV for large $c \tau$. 

\subsection{Extrapolation to High Luminosity LHC}
\label{subsec:HL-LHC}
While extrapolating phenomenological analyses from LHC data to larger luminosities is in principle a well understood task (taking into account some limitations in the publicly available information) and a straightforward procedure when prompt objects are involved, searches involving LLPs are quite a different case. The main reason is that the leading backgrounds involved are instrumental (interactions of particles with detector material, cosmic muons, beam halo, cavern radiation, detector noise, etc). These cannot be accurately simulated with Monte Carlo and state-of-the art public detector simulation packages and need to be taken from data. Indeed, every beam configuration would require a new estimation of these backgrounds. It is worth noting that usually an important bottleneck for these analyses is the poor trigger efficiency. Thus while the HL-LHC would be a much busier environment where the backgrounds would not be expected to scale with the luminosity, it is also likely that potential trigger upgrades could provide a larger signal acceptance, compensating for the larger rate of instrumental backgrounds.

We will follow here a conservative approach, where we extrapolate the current \emph{expected} number of background events in each search up to a total integrated luminosity of 3000 fb$^{-1}$.
Furthermore, we assume that the background (BG) uncertainties quoted by the collaborations for the 13 TeV analyses are dominated by systematics, so the relative BG uncertainty is assumed to remain constant when extrapolating to the HL-LHC:
\begin{equation}
    N_{BG}^{HL} = N_{BG}\frac{\mathcal{L}_{HL}}{\mathcal{L}} \text{ and } \delta_{BG}^{HL} =  \delta_{BG} \frac{N_{BG}^{HL}}{N_{BG}},
\end{equation}
where $N_{BG}$, $\delta_{BG}$ and $\mathcal{L}$ are the expected number of background events, its uncertainty and the integrated luminosity for the relevant Run II analyses, while
$N_{BG}^{HL}$, $\delta_{BG}^{HL}$ and $\mathcal{L}^{HL}$ are the corresponding values
for the high luminosity LHC.
In addition, in order to keep our extrapolation conservative,
we assume that the \emph{observed} number of events at the HL LHC 
is given by $N_{obs}^{HL} = N_{BG}^{HL} + 2\delta_{BG}^{HL}$.
This corresponds to the assumption that the HL analyses will measure a 
 $2\sigma$ upward fluctuation in all signal regions.
 As an additional layer of caution, we will not exploit the 14 TeV cross-sections and use the 13 TeV values.
 
\begin{figure}[!phtb]
\centering
\includegraphics[width=0.88\textwidth]{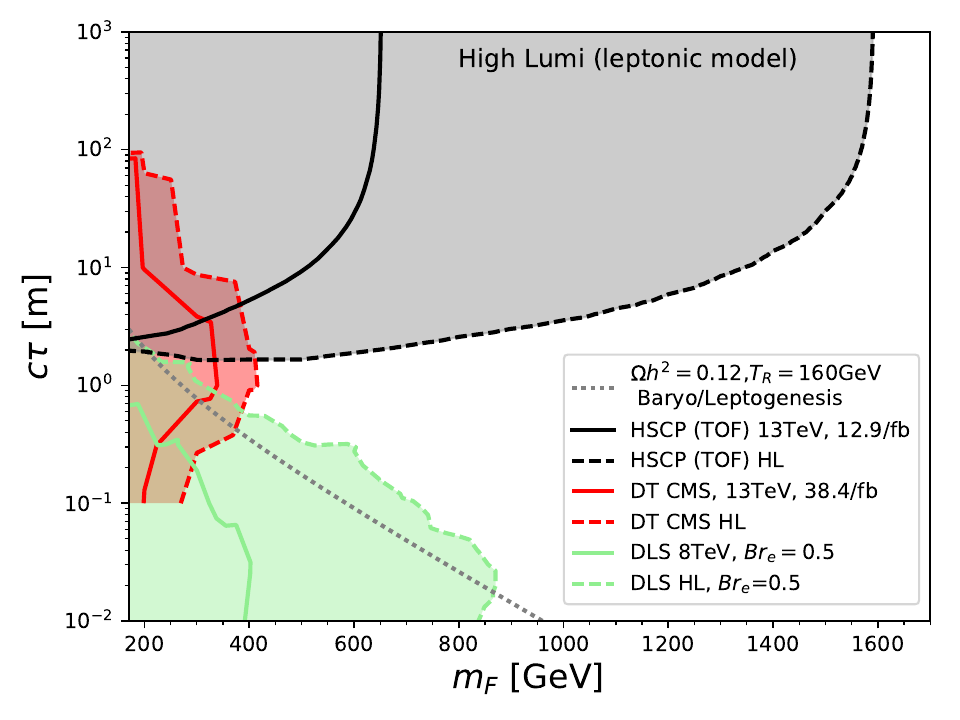}
\includegraphics[width=0.88\textwidth]{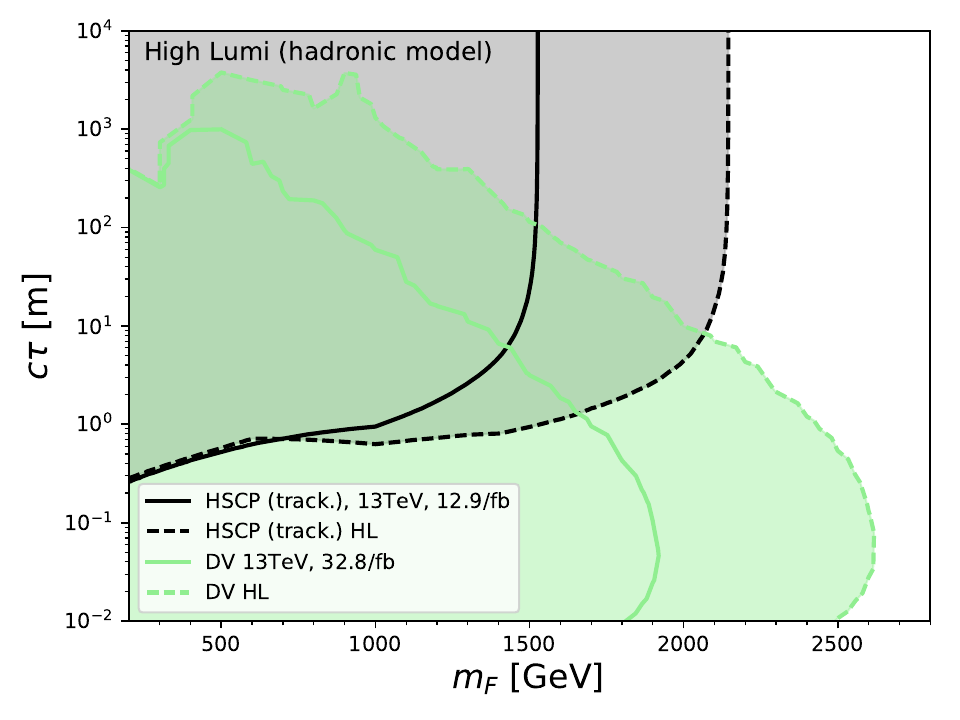}
\caption{Summary of the high luminosity LHC projections for the lepton-like (upper panel) and the quark-like (lower panel) FIMP scenarios. The lines correspond to contours of $\Omega_s h^2=0.12$ for the values of $m_s$ and $T_R$ given in the legend. See text for details.}
\label{fig:HL-projections}
\end{figure}

 Using the same recasting procedure described in Section~\ref{sec:collider}, but now assuming a $3000$ fb$^{-1}$ integrated luminosity and the extrapolation of background uncertainties and observed number of events described above, we re-compute the exclusion curves for the most
 constraining analyses from Section~\ref{sec:collider}.
 In Figure~\ref{fig:HL-projections} we compare the Run II (13 TeV) curves and the high luminosity (HL) projections for both the leptonic (top) and hadronic (bottom) models. 
 
 For the former, we expect the exclusion to extend to $m_{F} \simeq 1.5$~TeV
 for sufficiently large lifetimes via the HSCP searches, while the exclusion reaches only $m_F < 400$ GeV for $c \tau \in [0.6-2]$ m, and  $m_F \lesssim 800$ GeV for $c \tau \sim 1 $ cm. However the interesting result is that the area that would bring vanilla baryogenesis models in tension can now be tested mostly by the displaced lepton searches. In the case of the hadronic model, masses up to 2 TeV will be completely excluded, and for $c \tau \lesssim 1$ m the reach is extended up to $2.5-2.6$ TeV due to the DV+MET search, thus largely enlarging the coverage of the HL-LHC. 
 
\section{Conclusions}
\label{sec:conclu}
In this work we have analyzed the LHC prospects to probe freeze-in production of dark matter, in contrast to the standard \emph{WIMP-inspired} scenarios. To that extent we have first constructed a set of phenomenologically viable minimal models and we have then studied the current and expected constraints from a variety of long-lived particle (LLP) collider searches. We have made use of the existing searches for Heavy Stable Charged Particles (HSCP), Disappearing Tracks (DT), Displaced Leptons (DL) and Displaced vertices with transverse missing energy (DV+MET). Altogether these analyses probe complementary regions of the freeze-in parameter space, as their coverage is designed for specific lifetime ranges. 

The simplified models constructed rely on the existence of the $y Y X_{\rm} \chi$ operator. They feature two new dark particles (the dark matter candidate $\chi$ 
and the parent particle $Y$) and a SM particle $X_{\rm}$, with a coupling constant $y \sim 10^{-7} - 10^{-13}$ to ensure the freeze-in mechanism takes place. Under 
the simplistic assumption that $\chi$ is a pure SM singlet, $Y$ is forced to have the same gauge charges as the SM field, while its spin can be chosen as either fermion 
or vector. The whole set of quantum numbers of $Y$ determine the collider phenomenology: $Y$ is produced in pairs via gauge interactions as single production is 
proportional to $y$ and thus negligible. Out of the possible options, we have chosen a scalar dark matter particle and studied the cases where $X$ is a right-handed charged 
lepton and a right-handed quark. These Lagrangians have been implemented in  {\tt FeynRules} and the corresponding  {\tt UFO} and {\tt CalcHEP}  model files can be downloaded 
from~\cite{FICPLHC}.
 
The freeze-in mechanism relies not only on the $y$ coupling and the aforementioned operator to generate the correct relic abundance, but it also depends 
on the reheating temperature $T_R$, which sets the starting point for the freeze-in production of $Y$. Thus, a positive signal at colliders would allow us to extract 
a value of $T_R$ and thus directly compare to the critical temperature $T^*$ required for sphaleron transitions to be active. Finding $T_R < T^*$ would indeed exclude 
the simplest electroweak baryogenesis (and also leptogenesis) scenarios and eventually point towards super-cooled models. 

While the simplified models are subject to a plethora of constraints, including direct searches and indirect bounds such as electroweak precision data and flavor, 
most of these do not affect the parameter space as they are proportional to the freeze-in coupling $y$. Thus, the only collider bounds directly related to the 
gauge couplings come from the LEP bound on electroweak charged particles and the modified running of the strong coupling using LHC data, which are rather 
mild and constrain masses to be above a few hundred GeV~\cite{Becciolini:2014lya}. From the cosmological point of view, the only strong constraint is the 
lower bound on the DM mass,  $m_\chi \gtrsim$ 12 keV, coming from Lyman-$\alpha$ observations. Hence these freeze-in models are mainly constrained by long-lived 
particle searches at the LHC.

We have recasted all the aforementioned LLP searches, detailing in each case our procedure, which is widely applicable to all classes of LLP models. 
In particular, the HSCP searches target the $c \tau $ of the order of a meter to a few meters, while the DL, DV+MET and DT can cover lower 
lifetimes $ c \tau \lesssim 10$ cm. This simplistic view allows to foresee the existence of a difficult-to-probe region in the $m_{LLP}- c \tau $ plane 
at the interface between searches. We note however, that this region is more pronounced for the leptonic model, where masses above 250-300 GeV can not 
currently be constrained for $c \tau \in [0.3-2]$ m, while for the hadronic model one obtains a robust exclusion of LLP masses above 1.4 TeV. A naive 
and conservative extrapolation to HL-LHC still leaves a difficult region for $m_{LLP} \gtrsim $ 400 GeV for $c \tau \in [0.6-1.8]$ m and extends the exclusion 
in the hadronic model to 2 (2.5) TeV for long (short) lifetimes. We note that the LLP community is putting effort in designing new LLP strategies, 
triggers and detector upgrades, and thus significant improvements in the HL-LHC reach are a concrete possibility.

Finally, we would like to stress that we have studied here only the simplest possibilities, and that there is room for refinement 
not only from the model-building perspective (studying more complex scenarios, embedding these models into UV completions, examining alternative 
cosmological histories) but also from the collider studies (studying parameter extraction, designing new searches to cover the \emph{wedge}). If the 
absence of signals in direct and indirect detection experiments persists, this is a very promising avenue to explore in the future.

\acknowledgments 
We would like to thank Monika Blanke, Jared Evans, Enrique Fernandez-Martinez, Carlos Pena and Kalliopi Petraki for useful discussions. 
We also thank the 2017 ``Les Houches – Physics at TeV colliders"€ workshop where this work started.
A.L. was supported by the Sao Paulo Research Foundation (FAPESP), projects 2015/20570-1
and 2016/50338-6. A.G. and J.H. were supported by the Labex ILP (reference ANR-10-LABX-63) part of the Idex SUPER, and received financial state aid managed by the Agence Nationale de la Recherche, as part of the programme ``Investissements d'Avenir" under the reference ANR-11-IDEX-0004-02. J.H. was further supported by the DFG Emmy
Noether Grant No. HA 8555/1-1. D.S is supported by the National Science Foundation, U.S.A under Grant No. 1519045. 
S.S. was supported by the National Research Foundation of Korea (NRF), funded by the Ministry of Science \& ICT under contract NRF-2008-00460 and by the U.S. Department of Energy through the Distinguished Researcher Program from the Fermilab LHC Physics Center.
J.M.N. was supported by the Programa Atracci\'on de Talento de la Comunidad de Madrid via grant 2017-T1/TIC-5202, and also acknowledges support from the Spanish MINECO's ``Centro de Excelencia Severo Ochoa" Programme via grant SEV-2012-0249.
The work of A.P. and B.Z.  was partially supported by ``Investissements d'Avenir", Labex ENIGMASS, contract ANR-11-LABX-0012. A.P. acknowledges as well partial support from SOTON Diamond Jubilee Fellowship as well from Royal Society International Exchanges grant IE150682. B.Z. acknowledges also the Programa Atraccion de Talento de la Comunidad de Madrid under grant n.
2017-T2/TIC-5455.
N.D. was supported by  the OCEVU Labex (ANR-11-LABX-0060) as well as the A*MIDEX project (ANR-11-IDEX-0001-02) funded by the ``Investissements d'Avenir" French government program managed by the ANR.
The Feynman diagrams in this article have been drawn using Tikz-Feynman package~\cite{Ellis:2016jkw}.

\begin{appendix}

 \section{Rescaling HSCP Limits for Finite Lifetimes}
 \label{app:HSCP}

 As discussed in Sec.\ref{subsec:HSCP}, for scenarios where the $F$ lifetime is sufficiently small ($c \tau \lesssim$ few meters), the effective cross-section which must be considered when computing the constraints from HSCP searches is:
 \be
\sigma_{eff} = \sigma \times f_{LLP}(L,\tau),
\ee
where $\tau$ is the $F$ lifetime, $\sigma$ its total pair production cross-section and $L$ the relevant detector size.
Here we discuss how $f_{LLP}$ can be obtained, so we can properly apply the HSCP limits for the FIMP scenario with small lifetimes.

In order to determine $\sigma_{eff}$ (or equivalently $f_{LLP}$), we first compute how finite lifetimes affect the overall signal efficiency for a specific HSCP search, what requires a full recasting of the corresponding analysis.
Although recasting long-lived searches are usually extremely difficult, CMS has provided detailed efficiencies for the 8~TeV HSCP search from Ref.\cite{Chatrchyan:2013oca}.
In Ref.\cite{Khachatryan:2015lla} the online ($\epsilon^{on}$) and offline ($\epsilon^{off}$) selection efficiencies are given as a function of the truth level HSCP kinematics
and can be directly applied to hadron-level MC events 
without the need of a detector simulator (see Ref.\cite{Khachatryan:2015lla} for details).
For events with a single HSCP candidate, each event efficiency is simply given by:
\be
\epsilon (p) = \epsilon^{on}(p) \times \epsilon^{off}(p), \label{eq:eff1}
\ee
where $p$ is the HSCP 4-momentum and $\epsilon^{on/off}$ the CMS-provided efficiencies.
Since the CMS search requires at least one HSCP in each event, for events with two HSCP candidates the total event efficiency is:
\be
\epsilon_{12} = \epsilon_{12}^{on} \times \epsilon_{12}^{off}, \label{eq:eff12}
\ee
where
\be
\epsilon_{12}^{on/off} =  \epsilon_1^{on/off} \times \epsilon_2^{on/off} 
+ \epsilon_1 ^{on/off} \times \left( 1 - \epsilon_2^{on/off} \right) 
+ \epsilon_2^{on/off} \times \left( 1 - \epsilon_1^{on/off} \right) 
\ee
and $\epsilon_{i}^{on/off} = \epsilon^{on/off} (p_i)$ is the efficiency for the $i$-th HSCP in the event to pass the on/off-line selection.
The first term in the above expression corresponds to the efficiency for both HSCPs passing the selection, while
the last two terms correspond to the efficiency for only one HSCP candidate passing the selection.

Given the above expressions we can finally address how the 
finite lifetimes will affect the total signal efficiency.
In the FIMP models presented in Sec.\ref{subsec:models}, the HSCP candidates ($F$) are always pair produced, so initially we always have two HSCP candidates.
However, since the $F$ lifetime can be of the order of a meter or below, we must consider only the fraction of the produced $F$s which will decay after traversing a distance $L$ from the primary vertex.
As discussed in Sec.\ref{subsec:HSCP}, we take $L=3$~m for the tracker-only analysis and $L=11$~m for the tracker+TOF analysis.
The probability for the produced $F$ to cross a distance $L$ of the detector without decaying is given by:
\be
P(L,\tau,\beta) = e^{-L/\left(c \tau \gamma \beta \right)},
\ee
where $\tau$ is the $F$ proper lifetime, $\beta = v/c$ its boost and $\gamma = 1/\sqrt{1-\beta^2}$.
Therefore the expression in Eq.\ref{eq:eff12} must be modified in order to include the survival probability $P$:
\bea
\epsilon_{12} \to \epsilon(L,\tau) & = & 
P(L,\tau,\beta_1)\times P(L,\tau,\beta_2) \times \epsilon_{12}  \nonumber\\
& + & P(L,\tau,\beta_1)\times\left[1-P(L,\tau,\beta_2)\right] \times \epsilon_1 \nonumber\\ 
& + & P(L,\tau,\beta_2)\times\left[1-P(L,\tau,\beta_1)\right]\times \epsilon_2 \label{eq:reff}
\eea
with $\epsilon_{12}$ is given by Eq.\ref{eq:eff12} and 
$\epsilon_i \equiv \epsilon (p_i)$ is given by
Eq.\ref{eq:eff1}.
Once again the first term in the above expression corresponds to the probability for both $F$ particles decaying after the length $L$ and at least one of them passing the on- and off-line selection.
The last two terms correspond to the probability for a single $F$ decaying
after $L$ and satisfying the analysis selection.

Given $\epsilon(L,\tau)$, we can compute the overall signal efficiency for finite lifetimes simply adding the rescaled efficiency for each event:
\be
\epsilon_{signal}(L,\tau) = \frac{1}{N} \sum_{events} \epsilon(L,\tau)
\ee
Note that, for $\tau \to \infty$, we have $P(L,\tau,\beta) \to 1$ and $\epsilon(L,\tau) \to \epsilon_{12}$, as expected.
Finally, the effective fraction of long-lived particle relevant for the HSCP searches is given by:
\be
f_{LLP}(L,\tau) = \frac{\epsilon_{signal}(L,\tau)}{\epsilon_{signal}(L,\tau\to \infty)} = \sum_{events} \epsilon(L,\tau)/\sum_{events} \epsilon_{12} \label{eq:fLLP}
\ee
In order to explicitly compute $f_{LLP}$, we have generated 50k hadron-level MC events for pair production of $F$s using  {\sc MadGraph5\_aMC@NLO}~\cite{Alwall:2014hca} and Pythia 8~\cite{Sjostrand:2014zea}. Using the event kinematics we then computed the event efficiency for a given value of $\tau$ and $L$ using Eq.\ref{eq:reff}. Finally, using Eq.\ref{eq:fLLP}, we computed the effective fraction of long-lived particles which can be used to rescale the total $F$ production cross-section.

From the above discussion it is clear that a proper computation of the rescaling factor $f_{LLP}$ requires knowledge of the signal efficiencies ($\epsilon$). Although these efficiencies have been provided for the 8~TeV analysis for colour-neutral HSCPs in Ref.\cite{Khachatryan:2015lla}, the same is not true for the 13~TeV analysis and/or for {\it R-hadrons}.
However, since both analyses apply very similar selection cuts, we
assume that the 8~TeV efficiencies can also be (approximately) usedhttps://v2.overleaf.com/1589759815rdspcwwkzrxq
to compute $f_{LLP}$ for the 13~TeV search and the quark-like FIMP scenario. This is a good approximation, since, as shown in Eq.\ref{eq:fLLP}, $f_{LLP}$ is only affected by efficiency ratios, thus being fairly analysis independent.

\section{Calculation of upper limits }
\label{app:stat}
In this appendix we describe the adopted statistical analysis for the different searches we consider in this work.
\newline\newline\noindent
We assume that the likelihood of observing $n_i$ events in signal region $i$, given a hypothesis $\lambda_i=s_i+b_i$ is given by:
\be
L(n_i; s_i,b_i) = \prod_{i=\mathrm{SRs}} \,e^{-(s_i + b_i)}\, \frac{(s_i + b_i)^{n_i}}{n_i !} {\cal N}(b_i; \mu_{b,i},\sigma_{b,i})~,
\label{like}
\ee
i.e. the observed events follow as usual a Poisson distribution, and we model the background with a normal distribution with mean $\mu_{b,i}$ and standard deviation $\sigma_{b,i}$.
\newline\newline\noindent
We then adopt the -by now well established- $CL_s$ technique \cite{Read:2002hq} to extract the 95\%CL limits to our models from the different searches considered in this work, which effectively marginalises over the background uncertainties, and rescales the signal+background hypothesis's p-value, by the background-only p-value, to prevent from undesired strong signal exclusion limits in cases where the experiment's sensitivity is negligible. Note that this approach is not suitable for computing 2-sided intervals instead of upper limits. We have checked that, for cases of large number of observed events, the results obtained with this method are in good agreement with the method of profile likelihood, with likelihood-ratio as test-statistic.

\end{appendix}

\bibliographystyle{JHEP}
\bibliography{fimp}

\end{document}